\documentclass[onecolumn, 12pt]{IEEEtran}
\usepackage[utf8]{inputenc}
\usepackage{mathtools, bm, bbm, mathrsfs}
\usepackage{amssymb}
\usepackage{url, hyperref}
\hypersetup{
colorlinks=true,
linkcolor=black,
citecolor=black,
}
\usepackage{graphicx}
\usepackage{subfigure}
\usepackage[most]{tcolorbox}

\usepackage{xcolor}
\usepackage{float}
\usepackage{setspace}

\newcommand{\p}{\mathbbm{P}}
\newcommand{\mE}{\mathcal{E}}

\newcommand{\E}{\mathbbm{E}}
\newcommand{\Var}{\textsf{Var}}

\newcommand{\mA}{\mathcal{A}}

\newcommand{\Euav}{E_{\textsf{uav}}}

\usepackage{xcolor}
\usepackage{cite}
\allowdisplaybreaks
\usepackage{lmodern}
\usepackage{newtxtext}
\title{Lidar-Assisted Acquisition of Mobile Airborne FSO Terminals in a GPS-Denied Environment}
\author{Heyou Liu, Muhammad Salman Bashir, \IEEEmembership{Senior Member, IEEE}, Mohamed-Slim~Alouini,~\IEEEmembership{Fellow,~IEEE} \thanks{H.~Liu and M.-S.~Alouini are with the CEMSE Division, King Abdullah University of Science and Technology (KAUST), Thuwal 23955-6900, Kingdom of Saudi Arabia. M.~S.~Bashir is with the School of Computing and Engineering, University of Huddersfield, Huddersfield HD1 3DH United Kingdom. e-mail: (heyou.liu@kaust.edu.sa; m.bashir@hud.ac.uk; slim.alouini@kaust.edu.sa)}}
\date{February 2022}

\begin{document}

\maketitle

\begin{abstract}
    For acquisition of narrow-beam free-space optical (FSO) terminals, a Global Positioning System (GPS) is typically required for coarse localization of the terminal.  However, the GPS signal may be shadowed, or may not be present at all,  especially in rough or unnameable terrains.   In this study, we propose a lidar-assisted acquisition of an unmanned aerial vehicle (UAV) for FSO communications in a poor GPS environment. Such an acquisition system consists of a lidar subsystem and an FSO acquisition subsystem: The lidar subsystem is used for coarse acquisition  of the UAV, whereas, the FSO subsystem is utilized for fine acquisition to obtain the UAV's accurate position. This study investigates the optimal allocation of energy between the lidar and FSO subsystems to minimize the acquisition time. Here, we minimize the average acquisition time, and maximize the cumulative distribution function of acquisition time for a fixed threshold. We learn that an optimal value of the energy allocation factor exists that provides the best performance for the proposed acquisition system. 
\end{abstract}


\section{Introduction}
Because of the availability of sizable unoccupied bands in the optical domain of electromagnetic spectrum, free-space optics (FSO), or free-space laser communications, is expected to become a facilitator of high data-rates in the backhaul of future wireless communications systems  \cite{trichili2020roadmap}. Due to the fact that the path loss increases with longer wavelength, we have to squeeze transmitted laser energy into a narrow cone to be able to transmit optical signals over longer distances. Thus, FSO typically utilizes narrow transmission beams with a divergence angle that can be a small fraction of a milliradian (mrad) \cite{Bashir1, Bashir2}. However, communication with narrow beams requires sophisticated acquisition, tracking and pointing (ATP) subsystems \cite{nielsen1995pointing}.  Here, signal acquisition is the process of aligning the receiver in the arrival direction of the beam, whereas tracking is the maintenance of link alignment after the acquisition stage is complete \cite{kaymak2018survey}. Acquisition and tracking stages are especially important for establishing and maintaining optical links between mobile terminals such as unmanned aerial vehicles (UAV).



\subsection{Motivation of Current Study}
Unmanned aerial vehicles (UAVs) will continue to find major applications in 6G and beyond wireless communication systems \cite{Bashir:TCOM:2022,  Bashir:TWC:22}. A major application of UAVs involves provision of connectivity to far-flung/hard-to-reach areas where it will not be possible to connect the remote region with conventional tower-and-cable infrastructure. In the context of sensor networks, the UAVs may have to travel to a faraway  sensor network location, collect data, and return to base station for data offloading and processing. In this scenario, the UAV  deploys high-speed laser links in order to offload data as quickly as possible.   

 For the UAV to communicate with a ground station, it is important that the ground station acquires the UAV through a narrow-beam optical signal\footnote{We assume that the major computational resources lie at the ground station, and that the acquisition process takes place majorly at the ground station. Once the ground station acquires the UAV's angle-of-arrival, it forwards this information to the UAV on a low data-rate RF feedback channel. The UAV simply has to tilt its transmitter telescope in the direction of ground station to offload data.}. Typically, a GPS signal provides initial location coordinates to initiate the acquisition of a mobile UAV from the ground station \cite{xu2016gps}. However, it is essential to recognize that GPS signals cannot penetrate solid materials such as concrete, dense wood, or steel cladding as they have traveled approximately 12,500 miles through the atmosphere to reach the GPS receiver. In addition to this, an unamenable geographical terrain---such as thick forests, foliage, and mountains---can cause severe fading of the GPS signal on ground. Thus, when GPS fails, we have to use an alternative---such as a radar or a lidar instead of the GPS---to locate and acquire the UAV. 

\begin{figure}
     \centering
     \includegraphics[scale = 0.45]{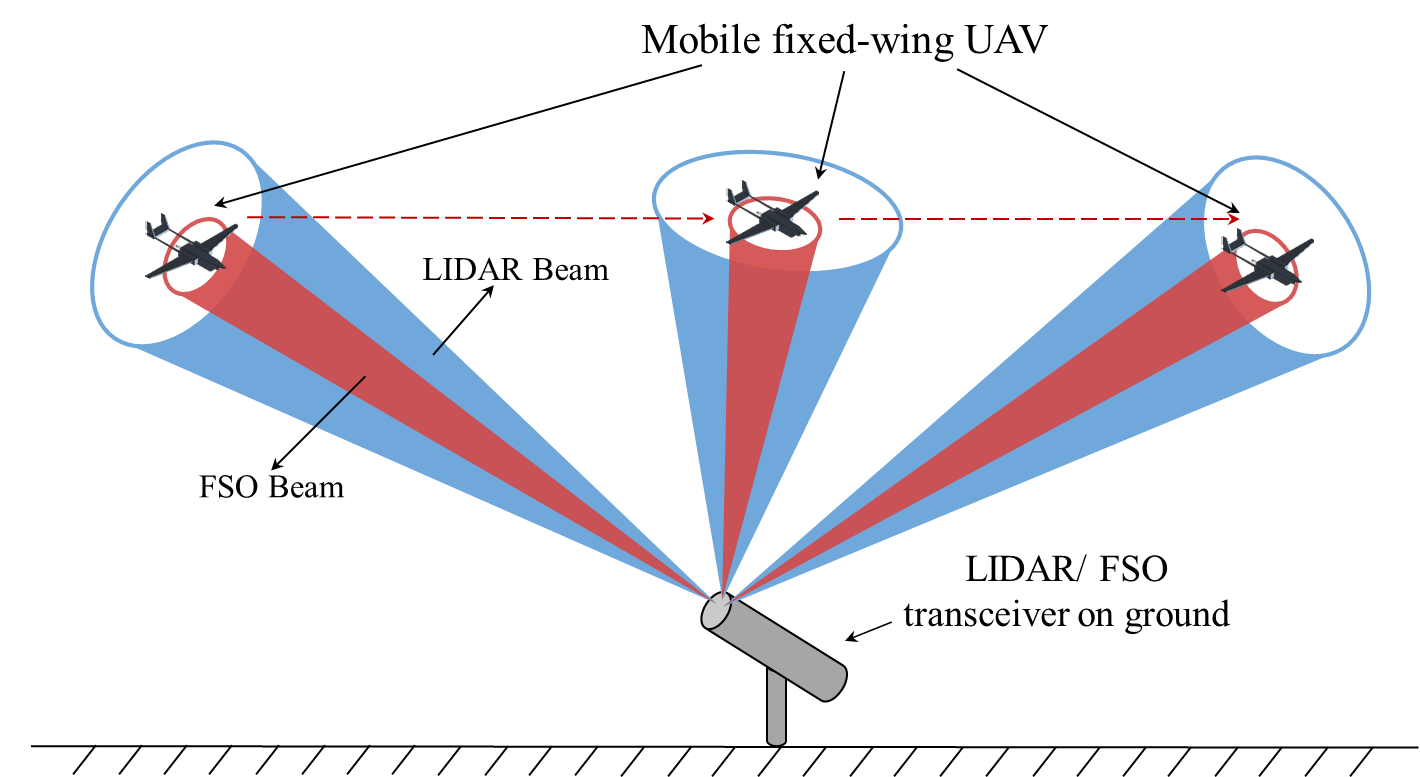}
     \caption{This figure depicts the dual lidar/FSO acquisition system trying to acquire a mobile UAV.}
     \label{Mode of acquisition system}
 \end{figure}

 For the purpose of acquisition, there are several advantages to using a lidar at the ground station instead of a radar: 1) Lidar is easy to integrate with existing FSO transceiver, 2) a lidar is better at detecting smaller UAVs, and 3) compared with the wider beamwidth of a radar, a lidar provides more security to avoid jamming and interference. In order to overcome the issue of GPS unavailability, we introduce the lidar-assisted acquisition of airborne terminals such as UAVs. For this case, a conceptual figure  is shown in Fig. \ref{Mode of acquisition system} in which a lidar/FSO transceiver on the ground is trying to acquire a mobile fixed-wing UAV by deploying  a wider lidar beam (blue) and a narrow FSO acquisition beam (red). Here, the lidar substitutes the GPS system in order to furnish rough location estimates of the mobile UAV to the FSO acquisition system in order to aid the acquisition process.


\subsection{Literature Review on Acquisition in Free-Space Optical Communications}

There are several studies in  literature that discuss beam acquisition and tracking for free-space optical terminals. For instance, the authors in \cite{lee2000pointing} proposed an acquisition technique in which the transmitter uses a narrow laser beam to search and locate the receiving terminal in an uncertainty region. In another study, the authors in \cite{li2011analytical} described a method to optimize the average acquisition time as a function of the uncertainty sphere angle for inter-satellite optical communications. However, the expressions of the probability of acquisition time are not derived. The authors in \cite{bashir2020signal} investigated the acquisition problem in free-space optical communications, determined the upper bounds on the mean acquisition time, minimized it numerically with respect to the beam radius, and derived the complementary distribution function of an upper bound on acquisition time in a closed-form. The study in \cite{harris2007minimization} presented an approach for minimizing optical link acquisition times by exploiting various wavelength-specific features. One article   \cite{Bashir:TCOM:21} considers adaptive acquisition schemes for photon-limited channels in free-space optical communications. The authors in  \cite{bashir2021optimal} analyzed the optimal power allocation between beam tracking and data detection channels of a single-detector optical receiver, and presented a comprehensive analysis of the error statistics of the centroid beam position estimator. Also, a few studies have concentrated on performance analysis of acquisition algorithms such as \cite{iftekharuddin1993acquisition,jono1999acquisition,ho2005performance}.

The is another body of literature that deals with the acquisition, tracking and pointing (ATP) of free space optical terminals such as \cite{kaymak2018survey}, \cite{carrasco2011low}, \cite{kim2007acquisition}, \cite{viswanath2015design}, and \cite{ho2007pointing}. The study in \cite{kaymak2018survey} gives a comprehensive survey on
acquisition, tracking and pointing (ATP) mechanisms, and discusses existing ATP mechanisms applied to free-space optical (FSO) communications systems. The authors in \cite{ho2007pointing} introduce an ATP mechanism that involves pointing the transmitter in the direction of a receiver, acquiring the incoming
light signal, and maintaining the FSO link by tracking the
position of a remote FSO terminal. The authors in \cite{kim2007acquisition} propose gimbal-less MEMS micro-mirrors for fast-tracking of time-varying beam position. However, the utilization of a wider beam may reduce the performance requirements of the employed ATP mechanism or allow for this mechanism's elimination \cite{kahn1997wireless,kaymak2017divergence}. It has been proposed that some FSO applications would benefit from adopting a wider beam to loosen the requirements of ATP mechanisms \cite{kotake2006new}. One study \cite{carrasco2011low} investigated an air-to-ground FSO communications system equipped with a MEMS retroreflector mirror as the primary component of an ATP mechanism for a UAV. The hybrid acquisition mechanism on the ground, comprised of a gimbal and FSM, provides both coarse and fine pointing. Furthermore, \cite{viswanath2015design} demonstrated a hybrid ATP mechanism with an array of MEMS mirrors and a motorized positioning system or a gimbal for a ground-to-satellite FSO communications system. The authors in \cite{rzasa2013pointing} explore the acquisition performance of a gimbal-based pointing system experimentally that uses spiral technique to scan the uncertainty region.

The authors in \cite{wang2002minimization} discuss a two-stage acquisition method for mobile FSO platforms. In that study, an array of detectors was utilized for the purpose of acquisition.  During the first stage, the terminals acquire each other's location; during the second, they verify each other’s identity through a code. The authors in \cite{arnon1997beam} and \cite{arnon2003optimization} presented some optimization models to provide the optimal beam radius for the minimum bit error rate (BER) and outage probability of an optical wireless link affected by pointing error and poor signal-to-noise ratio. 

\subsection{Contribution of This Study}
Even though the FSO literature contains numerous references on pointing, acquisition and tracking of optical beams, the study conducted in this paper is unique in that we tackled the acquisition problem for a mobile terminal in the absence of a GPS signal. As discussed earlier, the GPS is important to obtain the initial location coordinates of a mobile UAV, and depending on the accuracy of the GPS, the error in the location estimate can be large or small. Thus, based on the location estimate provided by the GPS, we may define a (spherical) region of uncertainty---also known as uncertainty or error sphere---around the estimated location of the target UAV. It is inside this uncertainty sphere that we have to search for the UAV using narrow FSO beams. In this study, the role of the GPS is taken over by a lidar in case the GPS signal is absent, and we examine the integration of a  lidar with a conventional FSO transmitter to obtain the initial region (or uncertainty sphere) of the target. Here, the lidar deploys a wider beam to localize the moving UAV and furnishes an uncertainty sphere at the location estimate of UAV. Thereafter, the FSO acquisition transmitter uses a super-thin beam to locate the UAV within the furnished uncertain region. One important aspect of the integration of lidar with FSO transmitter involves optimal energy allocation between the lidar and FSO channels in order to acquire the terminal as quickly as possible. 

In this study, we have conducted a detailed mathematical analysis to optimize the performance of a lidar-assisted acquisition system as a function of energy split factor $\alpha$. The quantity $\alpha$ is the fraction of transmitted energy going into lidar block, and $(1-\alpha)$ is reserved for FSO transmitter block. In this regard, we first derived the mean acquisition time and the complementary distribution function of the total acquisition time for the dual lidar-FSO transceiver system. As a next step, we optimized the energy allocation between the lidar and FSO acquisition subsystems to i) minimize the mean acquisition time and ii) maximize the CDF of acquisition time as a function of $\alpha$. Additionally, the mobility of the UAV has been taken into account by repeating the acquisition process at different points in time (and space) in case the acquisition attempt at a certain location fails. The number of failed acquisition attempts is modeled by a geometric random variable with a suitable mean.

\subsection{Organization of This Paper}
The structure of this paper is as follows: Section \ref{II} lays the foundation for this paper by presenting the system model of the acquisition problem with dual lidar-FSO transceiver.  Section \ref{acquisition haha} presents the acquisition algorithm  for the dual lidar/FSO transceiver. Section \ref{IV} analyzes the probability that the FSO acquisition system successfully detects the UAV for different levels of signal-to-noise ratio. Section \ref{V} undertakes  theoretical analysis on the effect of  energy split factor $\alpha$ over the average acquisition time. Here, the complementary cumulative distribution function of the acquisition time is also derived in closed-form. The optimization problem is also carried out in the same section. Section \ref{VI} discusses the minimization of  acquisition time for different conditions.
Section \ref{VII} summarizes the conclusions of this paper.

The list of important mathematical symbols in this paper is shown in Fig.~\ref{Table}.

 \begin{figure}
     \centering 
     \scalebox{1}{
     \begin{tabular}{{c}|{c}}
     \hline
     \hline
     \textbf{Symbol} & \textbf{Parameter}  \\ \hline
          $D$ & Distance between lidar and ground station\\ \hline
          $\alpha$ & the energy split factor \\ \hline
          $E_t$ & the total energy at the ground station \\ \hline
          $\theta_l$  & lidar beamwidth (half-angle) \\ \hline
          $\rho_l$ & lidar beam radius at distance $D$ \\ \hline
          $\theta_f$ & Tx beamwidth (half-angle)\\ \hline
          $\rho_f$ & Tx beam radius at distance $D$\\ \hline
          $\rho_{\textsf{uav}}$ & receiver aperture radius \\ \hline
          $(a_0,b_0)$ & Gaussian beam center \\ \hline
          $(u_0,v_0,z_0)$ & the location estimate of the UAV \\ \hline
          $\sigma$ & radar cross-section of the UAV \\ \hline
          $a_l$ & radius of lidar telescope \\ \hline 
          $\phi$ & azimuth angle \\ \hline
          $\psi$ & elevation angle\\ \hline
          $F$ & focal length of the lidar receiver telescope lens \\ \hline
          $\mathcal{A}$ & region of the detector array \\ \hline
          $\mE_\phi$ & azimuth angle estimation error\\ \hline
          $\mE_\psi$ & elevation angle estimation error\\ \hline
          $\sigma_s^2$ & variance of firing distribution\\ \hline
          $X$ & total number of acquisition attempts\\ \hline
          $p_X$ &(success) probability that receiver is located during one acquisition attempt \\ \hline
          $N$ & number of pulses fired during a successful acquisition attempt (truncated at $N_0$) \\ \hline
          $p_N$ & (success) probability that the receiver detects a pulse \\ \hline
          $N_0$ & maximum number of pulses fired during a given acquisition attempt \\ \hline
          $N'$ & untruncated version of $N$\\ \hline
          $\mathbbm{P}(\mathcal{C})$ & probability that receiver lies within Tx beam footprint\\ \hline
          $\mathbbm{P}(\mathcal{D}|\mathcal{C})$ & probability that receiver detects the laser pulse given the event $\mathcal{C}$ is true\\ \hline     
          \hline
     \end{tabular}
     }
     \caption{List of mathematical symbols used in this paper.}
     \label{Table}
 \end{figure}

 \section{System Model}\label{II}
\begin{figure}[ht]
    \centering
    \subfigure[Energy allocation on the ground station]{
    \label{Fig1.sub.a_first}
    \includegraphics[width=0.35\textwidth]{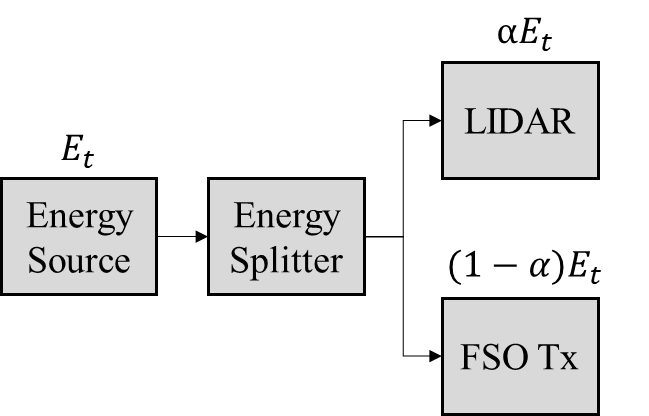}}
    \hspace{12mm}
    \subfigure[Model of the laser beam]{
    \label{Fig1.sub.b_second}
    \includegraphics[width=0.35\textwidth]{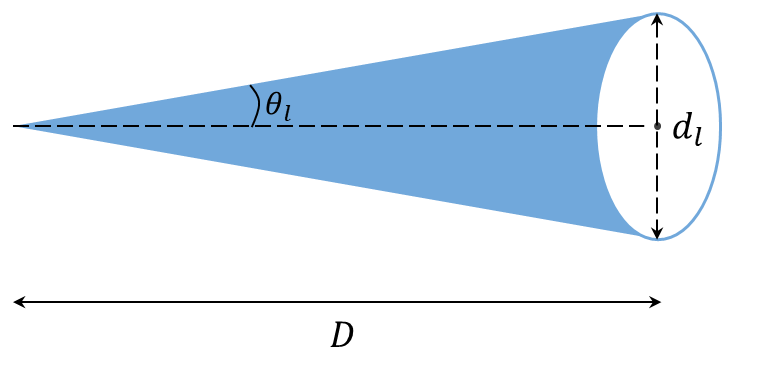}}
    \caption{A block diagram of energy division between the lidar and FSO acquisition subsystems (left) and the model of lidar beam (right).}
    \label{Fig1.main}
\end{figure}

For the purpose of UAV acquisition at the ground station, the total energy, $E_t$, is split between two blocks during the acquisition stage as shown in Fig. \ref{Fig1.sub.a_first}. The energy $\alpha E_t$ is reserved for the lidar, and the remaining energy $(1-\alpha)E_t$ is used for narrow-beam FSO acquisition transmitter.


As shown in Fig.~\ref{Fig1.main}, the (half-angle) beam divergence of the lidar is denoted by $\theta_l$. Let the diameter of the lidar beam at the target be denoted by $d_l$,  and the beam diameter of the FSO acquisition system be represented by $d_f$. The half-angle beam divergence of the FSO acquisition transmitter is denoted by $\theta_f$.
 


 
 \subsection{Lidar Ranging Equation}
 Let $(u_0, v_0, z_0)$ represents the location estimate of the UAV obtained with the help of the lidar. Here, the estimate $(u_0, v_0, z_0)$ denotes the center of uncertainty sphere, and this sphere will be scanned by the narrow-beam FSO acquisition transmitter. Since the height of the UAV is (approximately) known, we can reduce the three-dimensional problem to the two-dimensional case; henceforth, we call $(u_0, v_0)$ as the (two-dimensional) location estimate of the UAV.  
 
 As $(u_0, v_0)$ evolves in time due to mobility of UAV, the lidar updates the estimate of $(u_0, v_0)$ based on the angle-of-arrival of return signal. We assume that the transmit laser beam (for both lidar and FSO acquisition) has a Gaussian intensity profile:
 \begin{align}
     I_l(x,y) = \alpha E_t \frac{1}{2\pi \rho_l^2} \exp \left( - \frac{(x-a_0)^2 + (y-b_0)^2}{2\rho_l^2} \right), 
 \end{align}
 where $\rho_l = \frac{d_l}{2}$ is the lidar beam radius at the target UAV, $\alpha E_t$ is the total transmit energy of the lidar beam, and $(a_0, b_0)$ is the center of the Gaussian beam intensity on a two-dimensional plane $\mathcal{Z}$. The plane $\mathcal{Z}$ is parallel to the lidar transmit aperture plane and lies at a distance $D$ from the lidar (the quantity $D$ is the distance between the lidar and the UAV).
 
 In order to update $(u_0, v_0)$, we assume that the energy captured by the lidar transceiver---after reflection from the UAV--- is given by 
 \begin{align}
     E_r \coloneqq \frac{\alpha E_t}{2\pi \rho_l^2} \sigma \left( \frac{1}{4 \pi D^2} \right)^2 \pi a_l^2,
 \end{align}
 where $\sigma$ is the radar cross-section of the UAV, $a_l$ represents the radius of the lidar receiver telescope, and $\pi a_l^2$ represents the area of the lidar receiver aperture.  Since $\rho_l = \theta_l D$, we have that
 \begin{align}
     E_r &= \frac{\alpha E_t}{2\pi \theta_l^2 D^2} \sigma \left( \frac{1}{4 \pi D^2} \right)^2 \pi a_l^2 = \frac{\alpha E_t \sigma   a_l^2}{32\pi^2 \theta_l^2 D^6}. 
 \end{align}
 
 The amount of energy in a single photon is given by the Planck-Einstein relation $
     E_p = \frac{hc}{\lambda},
     $
 where $\lambda$ is the wavelength of light, $h$ is the Planck constant which is equal to $6.624 \times 10^{-34}$J/Hz, and $c$ is the speed of light in free-space: $c = 3 \times 10^8$ m/s. At $1550$ nm wavelength, the energy per photon $E_p = 3.9578 \times 10^{-18} \text{J}. $
 
 Thus, the average number of received signal photons is given by $
 \lambda_U \coloneqq \E[U] = \frac{E_r}{E_p},
 $
 where $U$ is the random number of photons captured by the lidar detector array (during the observation interval) after reflection from the target. The quanity $U$ is modeled by a Poisson distribution: $U\sim Pois(\lambda_U)$.

 \subsection{Estimation of Angle-of-Arrival with a Focal Plane Array}
 \begin{figure}
     \centering
     \includegraphics[scale=0.7]{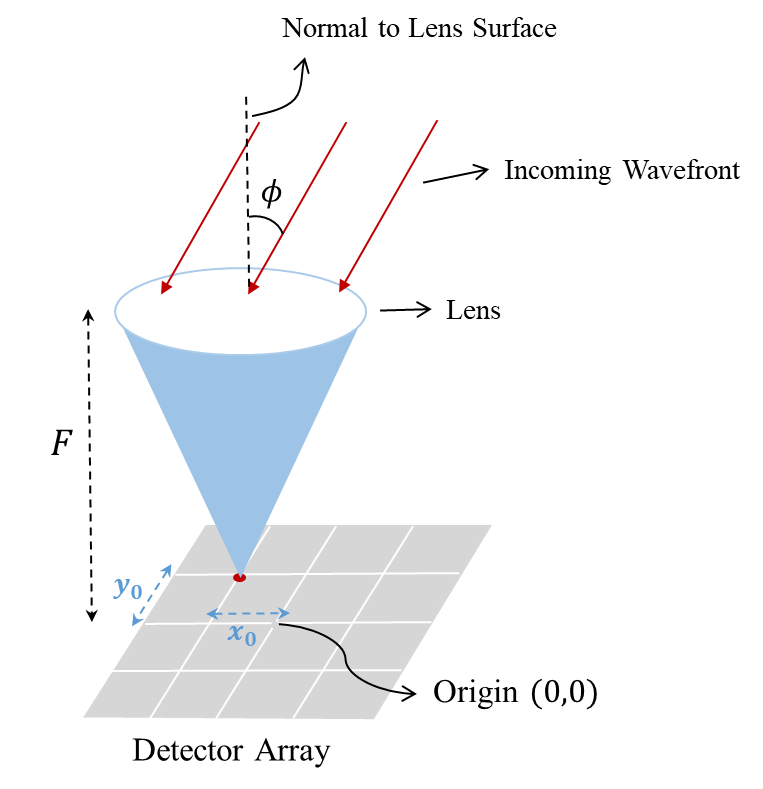}
     \caption{Projection of the return signal on the focal plane of lidar receiver lens.}
     \label{fig:lidar lens}
 \end{figure}
The energy captured by the lidar aperture is focused on an array of detectors which is placed in the focal plane of the lidar receiver telescope lens. The movement of the focused spot on the array indicates the changing angle-of-arrival of the reflected signal from the target as shown in Fig. \ref{fig:lidar lens}. Let the angle of arrival be denoted by $\phi$. Then, we have that the deviation of the spot (from the origin) on the array is given by 
\begin{align}
    x_0  \coloneqq F \sin(\phi), \quad
    y_0  \coloneqq F \sin(\varphi),
    \label{basic}
\end{align}
where $\phi$ represents the azimuth angle, $\varphi$ stands for the elevation angle, and $F$ represents the focal length of the lidar receiver telescope lens. Since the detector array is placed in the focal plane, the distance between the detector array and the lens is equal to $F$.

The intensity $\Lambda(x,y)$ of the beam spot on the array---measured in terms of average number of received photons---is modeled by a Gaussian distribution function as
\begin{align}
        \Lambda(x,y) = \frac{E_0}{2 \pi \rho^2} \exp\left( - \frac{(x-x_0)^2 + (y-y_0)^2}{2\rho^2} \right)\cdot \mathbbm{1}_\mA(x,y), 
\end{align}
where $(x_0, y_0)$ is the center of the spot on the array, $\rho$ is the effective spot radius and $\mA$ is the region of the detector array. The quantity $E_0$ is a normalization constant so that $\iint_\mA \Lambda(x, y) \, dx \, dy = \frac{E_r}{E_p}$. Here, we have assumed that the effect of background radiation and thermal noise is negligible, and any detected photon results only due to incident signal energy.

Let the estimate of $x_0$ be denoted by $\hat{x}_0$ and the estimate of $y_0$ be denoted by $\hat{y}_0$. Then, it is shown in \cite{bashir2021optimal} that the two-dimensional error vector 
\begin{align}\bm{\mE} \coloneqq \begin{bmatrix} \mE_x & \mE_y \end{bmatrix}^T \coloneqq \begin{bmatrix} x_0-\hat{x}_0 & y_0-\hat{y}_0 \end{bmatrix}^T,
\end{align}
is a zero-mean circularly symmetric Gaussian random vector for a continuous\footnote{All practical arrays are discrete (containing a finite number of elements) in nature. A continuous array is a limiting case of discrete array where the number of elements approach infinity for a constant array area. Thus, the area of each detector approaches zero for a continuous array. The continuous array assumption makes the problem simpler to analyze.} array and a centroid estimator of $(x_0, y_0)$.



 The location of the photons for a (continuous) photon counting detector array case is modeled by a Gaussian-Poisson process \cite{Snyder}. The centroid estimator of $(x_0, y_0)$ is
\begin{align}
    \hat{x}_0 \coloneqq \frac{1}{U}\sum_{i=1}^{U}x_i, \quad \, \,\hat{y}_0 \coloneqq \frac{1}{U}\sum_{i=1}^{U}y_i,
\end{align}
where $U$ is the number of the detected photons in some  observation interval, and $(x_i,y_i)$ is the random location of the $i$th photon. For two integers $i$, $j$ where $0 \leq i, j \leq U$, $x_i,x_j \sim \mathcal{N}(x_0, \rho^2)$ and $x_i \perp x_j$ for a continuous array. Assuming that the jitter is zero \cite{bashir2021optimal}, the mean-square error of the centroid estimator is
\begin{align}
    \textsf{MSE} &= \E\left[(x_0-\hat{x}_0)^2 + (y_0-\hat{y}_0)^2\right] = \E\left[(x_0-\hat{x}_0)^2\right] + \E\left[ (y_0-\hat{y}_0)^2\right].
    \label{variance of epsilon}
\end{align}
Here we first check whether the centroid estimator of $x_0$ is unbiased. The conditional expectation of $x_0$ is
\begin{align}
    \E\left[ \hat{x}_0 \, | \, U  \right] = \frac{1}{U} \sum_{i=1}^{U} \E \left[x_i \right] = \frac{1}{U} \cdot U x_0 = x_0.
\end{align}
Thus, the expectation of $\hat{x}_0$ is
    $\E\left[ \hat{x}_0 \right] = \E_{U}\left[ \E\left[ \hat{x}_0 \, | \, U  \right]\right] = x_0.$ 
Also, $\E\left[ \hat{y}_0 \right] = y_0$, which means the centroid estimator of $x_0$ is unbiased. Therefore, the MSE in (\ref{variance of epsilon}) can be simplified to
\begin{align}
    \textsf{MSE}
    &= \E\left[(\hat{x}_0 - \E\left[\hat{x}_0\right])^2\right] + \E\left[ (\hat{y}_0 - \E\left[\hat{y}_0\right])^2\right] = \Var\left[ \hat{x}_0 \right] + \Var\left[ \hat{y}_0 \right].
\end{align}
The conditional variance of  $\hat{x}_0$ is
\begin{align}
    \Var\left[ \hat{x}_0 \, | \, U \right] = \E\left[ \hat{x}_0^2 \, | \, U\right] -\left(\E\left[ \hat{x}_0 \, | \, U \right] \right)^2 = \E\left[ \hat{x}_0^2 \,|\,U \right] - x_0^2,
    \label{conditional variance}
\end{align}
where the second moment of $\hat{x}_0$ in (\ref{conditional variance}) is given by
\begin{align}
    \E\left[ \hat{x}_0^2 \, | \, U\right] &= \E\left[ \frac{1}{U^2} \left( \sum_{i=1}^{U}x_i \right)^2 \right] = \frac{1}{U^2} \E\left[ \sum_{i=1}^{U} x_i^2 + \sum_{i=1}^{U} \sum_{\substack{j=1 \\ j \neq i}}^{U} x_i x_j  \right] = \frac{1}{U^2} \left( \sum_{i=1}^{U} \E\left[ x_i^2 \right]  + \sum_{i=1}^{U} \sum_{\substack{j=1 \\ j \neq i}}^{U} \E[x_i x_j]  \right)
    \nonumber \\
    &= \frac{1}{U^2} \left( \sum_{i=1}^{U} \E\left[ x_i^2 \right]  + \sum_{i=1}^{U} \sum_{\substack{j=1 \\ j \neq i}}^{U} \E[x_i] \E[x_j]  \right) = \frac{1}{U^2}\left(  \sum_{i=1}^{U} \left( \Var[x_i] + \left( \E[x_i] \right)^2  \right) + \sum_{i=1}^{U} \sum_{\substack{j=1 \\ j \neq i}}^{U} \E[x_i] \E[x_j]\right) \nonumber\\
    &= \frac{1}{U^2} \left[ U\left(\rho^2+x_0^2 \right) + \left(U^2 - U\right) x_0^2  \right] \nonumber \\
    &= \frac{\rho^2}{U} + x_0^2.
\end{align}
Thus, the conditional variance of $x_0$ in (\ref{conditional variance}) can be simplified to:
\begin{align}
    \Var\left[\hat{x}_0 \, |\,U\right] = \E\left[ \hat{x}_0^2 \,|\,U \right] - x_0^2 = \frac{\rho^2}{U} +x_0^2 - x_0^2 = \frac{\rho^2}{U},
\end{align}
and the variance of $\hat{x}_0$ is obtained as
\begin{align}
    \Var[\hat{x}_0] = \sum_{u=0}^\infty \frac{\rho^2}{u}  \cdot \frac{\lambda_U^u}{u!} \,e^{-\lambda_U} \leq   \sigma^2_0 \, e^{-\lambda_U} + \sum_{u=1}^\infty \frac{\rho^2}{u} \cdot \frac{\lambda_U^u}{u!} \,e^{-\lambda_U},
    \label{x_0}
\end{align}
where $\sigma^2_0$ is the largest possible value of the estimate variance in a practical optical receiver. Let us adopt the convention that the beam center is located at the center of the array when $U$ is zero (that is, when no signal photon is detected). Therefore, for the $U=0$ case, the value of $\sigma^2_0$ is $\frac{|\mA|}{2}$. This is true because the value of the diagonal, denoted by $d$, of a square array is $d = \sqrt{2|\mA|}$, and the half-diagonal is $\frac{d}{2}= \frac{\sqrt{|\mA|}}{\sqrt{2}}$. Using our convention, $\frac{d}{2}$ is the largest possible Euclidean error between the actual beam center and the estimated beam center for $U=0$. Squaring the value of half-diagonal, we obtain $\sigma_0^2 = \frac{|\mA|}{2}$. For this scenario, the variance of $\hat{x}_0$ in (\ref{x_0}) is upper bounded by
\begin{align}
    \Var[\hat{x}_0] & \leq   \frac{|\mA|}{2}\,e^{-\lambda_U} + \sum_{u=1}^\infty \frac{\rho^2}{u}  \cdot \frac{\lambda_U^u}{u!} \,e^{-\lambda_U} = \frac{|\mA|}{2}\,e^{-\lambda_U} +\rho^2 e^{-\lambda_U}\left(-\gamma + \int_{-\infty}^{\lambda_U}\, \frac{e^t}{t}\,dt - \text{ln}(\lambda_U)\right) \nonumber \\
    &= \frac{|\mA|}{2}\,e^{-\lambda_U} +\rho^2 e^{-\lambda_U}\left(-\gamma + \textsf{Ei}(\lambda_U) - \ln (\lambda_U) \right), \label{ubound}
\end{align}
where $\textsf{Ei}(\lambda_U)$ is the exponential integral function, and $\gamma$ is the Euler's constant which is approximately 0.577216. It can be shown that the upper bound in \eqref{ubound} is a positive quantity (see Appendix).

Due to the circularly symmetric nature of the Gaussian beam and the approximation of the practical  detector array with a continuous array, we have that $\Var[\hat{y}_0] = \Var[\hat{x}_0]$.
Since the estimation errors along $x$ and $y$ dimensions are independent due to the circularly symmetric Gaussian beam, we will analyze the (one-dimensional) error in $\phi$ only since the same analysis will hold for the estimation error in $\varphi$. 

A straightforward estimate of the angle-of-arrival $\phi$ is
\begin{align}
    \hat{\phi} \coloneqq \sin^{-1} \left( \frac{\hat{x}_0}{F} \right).
    \label{related}
\end{align}
However, it is hard to deduce an obvious relationship between $\hat{\phi}$ and $\hat{x}_0$ in \eqref{related}; therefore we can consider it from a different perspective. We have that $\hat{\phi} = \phi + \mE_\phi$ (small error $(x_0 - \hat{x}_0)$ leads to the error  $(\phi - \hat{\phi})$). Then, 
\begin{align} 
&\hat{x}_0 = F \sin \left( \hat{\phi} \right) \implies x_0 + \mE_x = F \sin \left( \phi + \mE_\phi \right) = F \left(\sin(\phi) \cos (\mE_\phi) + \cos(\phi) \sin(\mE_\phi) \right),
\label{wwwww}
\end{align}
since  $\mE_\phi$ is small enough for most practical systems such that we can make approximations $\sin(\mE_\phi)\approx\mE_\phi$ and  $\cos(\mE_\phi) \approx 1$. Then,  \eqref{wwwww} can be simplified to
\begin{align}
    x_0 + \mE_x \approx F \sin(\phi)  + F \cos(\phi) \mE_\phi,
\end{align}
where $x_0 = F \sin (\phi)$ can be obtained from \eqref{basic}. Therefore,  we have that
\begin{align}
    \mE_x = F \cos(\phi) \mE_\phi \implies \mE_\phi = \frac{\mE_x}{F \cos(\phi)}.
\end{align}
Thus, $\mE_\phi$ is a zero-mean Gaussian random variable since $\mE_x\sim \mathcal{N}(0,\sigma_{\hat{x}_0}^2)$, and thus its variance \begin{align}
\Var[\mE_\phi]& = \left(\frac{1}{F\cos(\phi)}\right)^2 \Var[\mE_x]= \left(\frac{1}{F\cos(\phi)}\right)^2 \sigma_{\hat{x}_0}^2 \nonumber \\
& \leq  \left(\frac{1}{F\cos(\phi)}\right)^2 \,e^{-\lambda_U}\, \left( \frac{|\mA|}{2} - \rho^2 \left( -\textsf{Ei}(\lambda_U) + \ln (\lambda_U) + \gamma \right) \right). \label{phi}
\end{align}
Similarly, it can be shown that  
\begin{align}
\Var[\mE_\varphi] \leq  \left(\frac{1}{F\cos(\varphi)}\right)^2 \,e^{-\lambda_U}\, \left( \frac{|\mA|}{2} - \rho^2 \left( -\textsf{Ei}(\lambda_U) + \ln (\lambda_U) + \gamma \right) \right). \label{psi}
\end{align}

\subsection{Volume and Shape of Uncertainty Sphere}

The angular errors $\mE_\phi$ and $\mE_\varphi$ represent the lidar measurement error in the angle-of-arrival. Thus, the volume of uncertainty sphere is determined by the errors $\mE_\phi$ and $\mE_\varphi$. 

The shape of the uncertainty sphere depends on the variance of the elevation angle and azimuth angle---that is, the relationship between $\phi$ and $\varphi$. Based on $\phi$ and $\varphi$, we deal with the following two scenarios:
\subsubsection{Circular Uncertainty Sphere}
The assumption $\phi \approx \varphi$ implies $\Var[\mE_{\phi}] \approx \Var[\mE_{\varphi}]$ (please see \eqref{phi} and \eqref{psi}), which makes the shape of the uncertainty sphere circular. { In this case, the volume of uncertainty sphere (circular sphere) is given by $\pi \mathcal{R}_u^2$ where 
\begin{align}
\mathcal{R}_u \coloneqq 3 \sqrt{\Var[\mE]}D, \label{true_radius}
\end{align}
where $\mathcal{R}_u$ is defined as the radius of uncertainty sphere. 
(please refer to Fig \ref{update}) where $\mE$ is $\Var[\mE] \coloneqq {\Var[\mE_\phi]} ={\Var[\mE_\varphi]}$. Here we use ``3-sigma rule" of the normal distribution to approximate $\mathcal{R}_u$.}  Let $\sigma_\mE^2$ be an upper bound on $\Var[\mE]$. We then have that
\begin{align}
    \Var[\mE] \leq \sigma_\mE^2 \coloneqq  \left(\frac{1}{F\cos(\varphi)}\right)^2 \,e^{-\lambda_U}\, \left( \frac{|\mA|}{2} - \rho^2 \left( -\textsf{Ei}(\lambda_U) + \ln (\lambda_U) + \gamma \right) \right). \label{variance of Uncertainty Sphere}
\end{align}

\subsubsection{Elliptical Sphere}
In this case $\phi \neq \varphi$, which implies 
$\Var[\mE_{\phi}] \neq \Var[\mE_{\varphi}] $. This leads to an  elliptical uncertainty sphere, where the semi-major axis and semi-minor axis are $R_a \approx 3\sigma_{\mE_{\phi}}D$, $R_b \approx 3\sigma_{\mE_{\varphi}}D$, respectively. In this case, the variance of estimation error alone $x$ axis and $y$ axis are $\sigma_{\mE_{\phi}}, \sigma_{\mE_{\varphi}}$, respectively:
\begin{align}
      \Var[\mE_\phi] \leq  \sigma_{\mE_{\phi}}^2 \coloneqq  \left(\frac{1}{F\cos(\phi)}\right)^2 \,e^{-\lambda_U}\, \left( \frac{|\mA|}{2} - \rho^2 \left( -\textsf{Ei}(\lambda_U) + \ln (\lambda_U) + \gamma \right) \right),  \\
     \Var[\mE_\varphi] \leq \sigma_{\mE_{\varphi}}^2 \coloneqq  \left(\frac{1}{F\cos(\varphi)}\right)^2 \,e^{-\lambda_U}\, \left( \frac{|\mA|}{2} - \rho^2 \left( -\textsf{Ei}(\lambda_U) + \ln (\lambda_U) + \gamma \right) \right),
\end{align}
where $\sigma_{\mE_{\phi}}^2$ and $\sigma_{\mE_{\varphi}}^2$ are upper bounds on the actual variance of $\mE_\phi$ and $\mE_\varphi$, respectively. 

As discussed before, the uncertainty sphere represents the measurement error with a lidar (where the measurement error is approximately Gaussian for centroid estimator and continuous arrays). Hence, we note that the ``true'' location of UAV inside the uncertainty sphere is a zero mean Gaussian random vector with a  covariance matrix
$\mathbf{C}_{\phi, \varphi} \coloneqq \begin{bmatrix}
    \Var[\mE_\phi] & 0 \\
    0 & \Var[\mE_\varphi]
\end{bmatrix}$. This covariance matrix is upper bounded by
\begin{align}
\mathbf{C}_{\phi, \varphi} \leq \mathbf{C} \coloneqq  \begin{bmatrix} 
\sigma_{\mE_\phi}^2 & 0 \\ 
0 & \sigma_{\mE_\varphi}^2  
\end{bmatrix}, \label{uav_location}
\end{align}
where the eigenvalues of $\mathbf{C}$ represent an upper bound on the eigenvalues of $\mathbf{C}_{\phi, \varphi}$. { From this point onward, we will use the variable $R_u$ as the radius of uncertainty sphere, where $R_u$ is defined as $R_u \coloneqq 3 \sigma_\mE D$, and the square of $\sigma_\mE$ is defined by \eqref{variance of Uncertainty Sphere}. So even though the proposed radius $R_u$ is greater than the actual radius $\mathcal{R}_u$ defined in \eqref{true_radius}, the optimization problem and results do not change whether $R_u$ or $\mathcal{R}_u$ is employed to represent the radius of uncertainty sphere. }

\subsection{The Shotgun Approach}

Once the lidar has established an uncertainty sphere in space, the FSO acquisition system begins to search/scan the sphere to locate the UAV. In the shotgun approach, the FSO acquisition system randomly scans the uncertainty region by firing pulses at random points inside the uncertainty region \cite{Bashir:TCOM:21}. According to the shape of the uncertainty sphere, the probability that the UAV detects the pulse from ground station is discussed for the following two scenarios:
\subsubsection{Circular Uncertainty Sphere} 
 In this case, the locations of the pulses fired towards the uncertainty region are chosen randomly from a zero-mean circular symmetric Gaussian distribution (also known as the \emph{firing distribution}). This Gaussian distribution is represented by $\mathcal{N} \left(0, \sigma_s^2 \mathbf{I}_2 \right)$ where $\mathbf{I}_2$ is a $2 \times 2$ identity matrix and $\sigma_s^2 \coloneqq 2 \sigma_\mE^2 D^2$.
The distribution of the pulse is therefore given by
\begin{align}
    f(x, y) = \frac{1}{2\pi \sigma_s^{2}}\exp \left (\frac{x^2+y^2}{2\sigma_s^{2}} \right). \label{circ_fire_dist}
\end{align}

\subsubsection{Elliptical Sphere}

In this case, the locations of these pulses are chosen randomly from a zero-mean elliptical Gaussian firing distribution, where the variance along $x$ and $y$ axes are independent but not equal: 
\begin{align}
    f(x, y) = \frac{1}{2\pi\sigma_{s1}\sigma_{s2}}\exp \left( -\frac{1}{2} \left( \frac{x^2}{\sigma_{s1}^2} + \frac{y^2}{\sigma_{s2}^2} \right)  \right).
\end{align}
where $\sigma_{s1}^2 \coloneqq  \sigma_{\mE_{\phi}}^2 D^2$ and $\sigma_{s2}^2 \coloneqq  \sigma_{\mE_{\varphi}}^2 D^2$.


\section{Acquisition Algorithm and Acquisition Time}\label{acquisition haha}

In this section, we discuss the derivation of acquisition time expressions. Since the acquisition time depends on the algorithm used for the purpose of acquisition, we first define the algorithm for the dual lidar-FSO based acquisition system. 
\subsection{Acquisition Algorithm}
\begin{figure}
    \centering
    \includegraphics[scale=0.45]{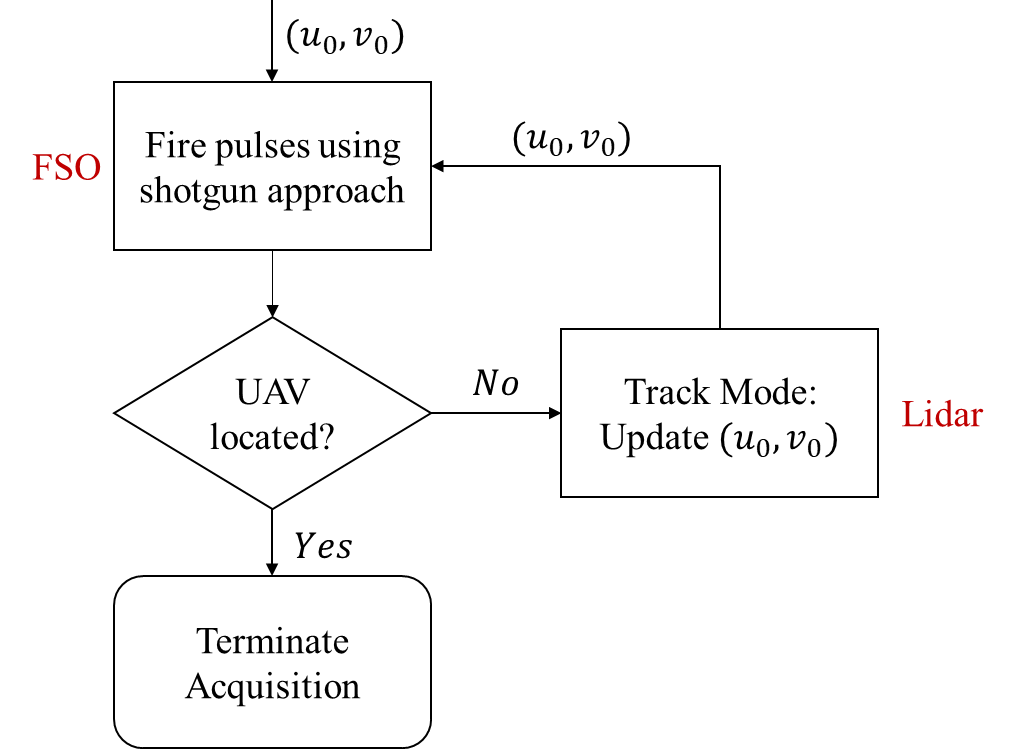}
    \caption{Flowchart of the lidar-assisted acquisition technique.}
    \label{flow}
\end{figure}

\begin{figure}
    \centering
    \includegraphics[scale=0.45]{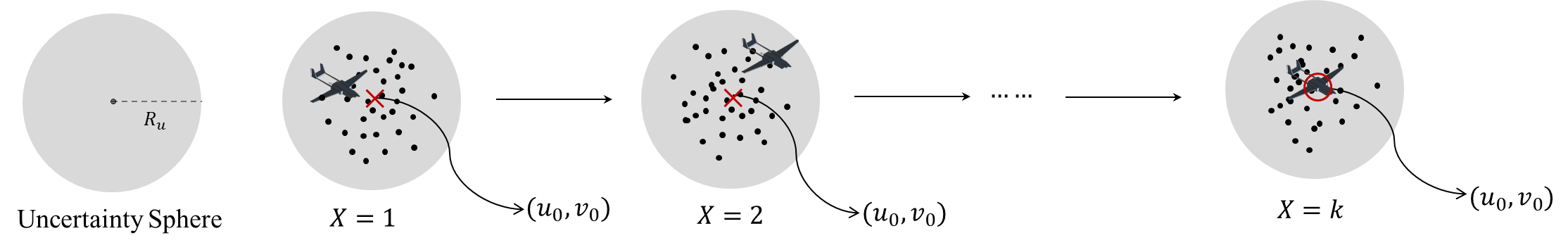}
    \caption{The left figure shows the uncertainty sphere with radius $R_u$.  The right figure shows how the lidar updates the uncertainty region center $(u_0, v_0)$ during tracking stage. The black dots indicates pulses of the firing distribution.}
    \label{update}
\end{figure}

We propose the following algorithm for the lidar-assisted acquisition scheme: 
\begin{enumerate}
    \item The lidar gives us an initial estimate of UAV coordinates $(u_0, v_0)$ based on the return signal from UAV.
    \item The FSO acquisition transmitter searches the uncertainty region with the center $(u_0, v_0)$ by firing pulses using the shotgun approach. A maximum of $N_0$ pulses are fired to locate the UAV. The acquisition algorithm stops if the target is located within $N < N_0$ pulse. Else, we move to  Step~3.
    \item { If the UAV is not located during Step~2, the lidar transmits another pulse in the direction of $(u_0, v_0)$, and based on the return signal, it updates the estimate of the target location $(u_0, v_0)$\footnote{Here, the lidar beam is wide enough so that even though the lidar pulse is fired in the original direction $(u_0, v_0)$, the mobile UAV is still within the lidar's FOV and generates a sufficiently large return signal to update $(u_0, v_0)$.}, as depicted in Fig. \ref{update}. We then go back to Step~2 to search the new uncertainty sphere with updated center $(u_0, v_0)$.}
\end{enumerate}
The  flowchart of this algorithm is described in Fig.~\ref{flow}.

We assume that the UAV is in the lidar's field-of-view when the acquisition process begins\footnote{Strictly speaking, the lidar is in the ``search'' mode initially when the UAV is not within its field-of-view. Once the UAV is within the field-of-view, the lidar enters the ``track'' mode.}. In the track\footnote{Here we use the word ``tracking'' to represent the estimation of parameter $(u_0, v_0)$ based on 
 only the current set of observations or measurements. This is in contrast with Bayesian tracking where the observations from the present as well as the past are fused to refine the estimate. } mode of the lidar,  let $T_1$ (a deterministic quantity) be the round-trip time of the pulse and the time taken to detect and process the reflected signal by the lidar.  Let $T_2$ be the time interval between each pulse of the shotgun technique\footnote{The FSO transmitter has to wait a certain amount of time before it fires the next pulse to locate the receiver. This waiting time is denoted by $T_2$. During this waiting period, the FSO transmitter listens for an acknowledgement from the UAV receiver. If the receiver detects the pulse, it informs the FSO transmitter to terminate the acquisition process. Otherwise, the acquisition process continues. }. 

We define an \emph{acquisition attempt} as one attempt to acquire the UAV. One acquisition attempt involves firing one pulse of the lidar followed by $N_0$ pulses fired by the FSO acquisition system. If one acquisition attempt fails, we carry out the next acquisition attempt until we locate the UAV. For this scenario, the total acquisition time is given by
\begin{equation}
    T \coloneqq  X T_1 + (X-1) N_0 T_2 + N T_2, \label{total acquisition time}
\end{equation}
where the random variable $X$ is the total number of acquisition attempts, and the quantity $N$ is the number of pulses fired by FSO acquisition system during the (final) successful acquisition attempt ($N<N_0$). 

\subsection{Expectation of $X$ and $N$}
In order to compute the expected value of acquisition time, we need to find the expectation of $X$ and $N$. The quantity $X$ is a geometric random variable with success probability $p_X$, and the quantity $N$ is a truncated geometric random variable with success probability $p_N$. Both $X$ and $N$ are treated as independent random variables. The pmf  of $X$ is given by
\begin{align}
    \p\left( \left\{X = k\right\} \right) = (1-p_X)^{k-1} p_X, \quad k = 1, 2, \dotsc
\end{align}
where the integer $k$ represents the total number of trials ($k-1$ failures to locate the terminal and the final attempt in which the terminal is located successfully). The pmf of $N$ is given by
\begin{align}
  \p(\left\{ N = k \right\}) = \frac{(1-p_N)^{k-1} p_N}{F_{N'}(N_0) - F_{N'}(1)}, \quad k=1, 2, \dotsc, N_0,
\end{align}
where $F_{N'}(\cdot)$ is the cumulative distribution function of the (untruncated) geometric random variable $N'$ whose PMF and CDF are given by $\p(\{N' = k\}) \coloneqq (1-p_N)^{k-1} p_N$ for $ k = 1, 2, \dotsc$, and $F_{N'}(x) \coloneqq 1-(1-p_N)^{\lfloor x \rfloor}$ for $x \geq 0$, respectively. The quantity $p_N$ is the probability that the laser shines the receiver aperture with a pulse, and the receiver detects the pulse successfully. We will discuss the derivation of $p_N$ in Section~{\ref{IV}}.

The expectation of X is given by
\begin{align}
    \E[X] = \dfrac{1}{p_X},
\end{align}
and the expectation of $N$ is given by
\begin{align}
    \E[N]
    &= \frac{1}{[F_{N'}(N_0) - F_{N'}(1)]}\cdot \sum_{k=1}^{N_0} k (1-p_N)^{k-1} \cdot p_N   \nonumber \\
    &= \frac{1}{[F_{N'}(N_0) - F_{N'}(1)]}\cdot  p_N  \sum_{k=1}^{N_0} k (1-p_N)^{k-1}. 
\end{align}
In order to get a closed-form expression of $\E[N]$, let $S = \sum_{k=1}^{N_0} k (1-p_N)^{k-1}.$ Then,  $(1-p_N) \cdot S = (1-p_N) + 2\cdot (1-p_N)^2 + 3\cdot (1-p_N)^3 + \dots + (N_0-1) \cdot (1-p_N)^{N_0-1} + N_0 \cdot (1-p_N)^{N_0}$. The difference
\begin{align}
    S - (1-p_N) \cdot S &= 1 + (1-p_N) + (1-p_N)^2 + \dots + (1-p_N)^{N_0-1} - N_0 \cdot (1-p_N)^{N_0} \nonumber\\
    \implies p_N \cdot S &= \frac{1-(1-p_N)^{N_0}}{1-(1-p_N)} - N_0\cdot(1-p_N)^{N_0} \nonumber \\
    \implies S &=  \frac{1-(1-p_N)^{N_0} \cdot (1+N_0 p_N)}{p_N^2}.
\end{align}
Thus, 
\begin{align}
    \E[N] &= \frac{1}{[F_{N'}(N_0) - F_{N'}(1)]} \cdot p_N \cdot S = \frac{1}{[F_{N'}(N_0) - F_{N'}(1)]} \cdot p_N \cdot \frac{1-(1-p_N)^{N_0} \cdot (1+N_0 p_N)}{p_N^2} \nonumber\\
    & =  \frac{1}{[(1-p_N)-(1-p_N)^{N_0} ]} \left(\frac{1-(1-p_N)^{N_0}(1+N_0 p_N)}{p_N} \right).
\end{align}

\subsection{Expression of $p_X$}
The probability $p_X$ can be obtained from $p_N$, since $1-p_X$ represents the probability of the event that the receiver is not located in an acquisition attempt by firing $N_0$ pulses, whereas $(1-p_N)$ is the probability that the receiver does not detect a firing pulse during the acquisition attempt. Therefore, the relationship between them is given by
\begin{align}
    &1-p_X = (1-p_N)^{N_0} \implies p_X = 1 - (1-p_N)^{N_0}, \label{relationship between p_X and p_N} 
\end{align}
and, thus, the expectation of $X$ is given by
\begin{align}
    &\E[X] = \frac{1}{1-(1-p_N)^{N_0}}.
\end{align}

\section{Derivation of $p_N$}\label{IV}
The event that the UAV receiver reports a pulse detection can be expressed as the intersection of two independent events: $A$ and $B$.  The event $A$ occurs when the UAV's aperture lies inside the beam footprint, and $B$ is the event that the UAV receiver reports a pulse detection. Thus, we have that
\begin{align}
    p_N \coloneqq \mathbbm{P}(\mathcal{C})  \mathbbm{P}(\mathcal{D}|\mathcal{C}), \label{P_N}
\end{align}
where $\mathbbm{P}(\mathcal{C})$ is the \emph{coverage} or \emph{overlap probability}---the probability that the laser pulse covers or shines the receiver. The measure $\mathbbm{P}(\mathcal{D}|\mathcal{C})$ represents a \emph{conditional detection probability} which indicates that the UAV receiver is able to detect a given laser pulse given that event $\mathcal{C}$ is true.

\subsection{Derivation of Coverage Probability $\mathbbm{P}(\mathcal{C})$}




1) \emph{Circular Sphere Case (Rayleigh Distribution)}

Let the random location of the FSO acquisition pulse be $\bm{X} = (X, Y)$ and the random location of the UAV receiver inside the uncertainty sphere be $\bm{X_0}=(X_0, Y_0)$. Through \eqref{circ_fire_dist}, we know that $\bm{X} \sim \mathcal{N}\left( \bm{0}, \sigma_s^2 \mathbf{I}_2 \right)$, and through \eqref{uav_location}, we note that $\bm{X}_0 \sim \mathcal{N}\left(\bm{0}, \sigma_\mE^2 D^2 \mathbf{I}_2 \right)$. Additionally, we find that the beam's footprint completely covers the location of the receiver aperture---that is the event $\mathcal{C}$ is true---when the distance between the center of the pulse and the receiver is less than the difference of their aperture radii:
\begin{align}
    \| \bm{X} - \bm{X}_0 \|_2 = \sqrt{(X - X_0)^2 + (Y-Y_0)^2} < \rho_f -\rho_{\textsf{uav}},
\end{align}
where the quantity $(X-X_0) \sim \mathcal{N}(0, \sigma_s^2 +  \sigma_\mE^2 D^2)$, and $ (Y-Y_0) \sim \mathcal{N}(0, \sigma_s^2+ \sigma_\mE^2 D^2)$. Since $X-X_0 \perp Y-Y_0$, we have that $\| \bm{X} - \bm{X}_0 \|_2$ is a Rayleigh random variable with scale parameter $\sqrt{ \sigma_s^2+ \sigma_\mE^2 D^2}$. In this case,  the coverage probability is given by
\begin{align}
    \mathbbm{P}(\mathcal{C}) &\coloneqq \p(\{\| \bm{X} - \bm{X}_0 \|_2 < \rho_f -\rho_{\textsf{uav}} \}) \nonumber \\
    & = 1 - \exp \left( -\frac{\left(\rho_f - \rho_{\textsf{uav}} \right)^2}{ 2 \left( \sigma_s^2 + \sigma_\mE^2 D^2\right)} \right).  \label{p_C}
\end{align}

2) \emph{Elliptical Sphere Case (Hoyt Distribution)}

In this more general scenario, the variance of the random location of the pulse and the the variance of UAV receiver along each of the $x$ and $y$ axes are not necessarily the same. Thus, we have that $(X-X_0) \sim \mathcal{N}(0,\sigma_{s1}^2 + \sigma_{\mE_{\phi}}^2D^2)$, $(Y-Y_0)\sim \mathcal{N}(0,\sigma_{s2}^2 + \sigma_{\mE_{\varphi}}^2D^2)$, which means that  $\|\bm{X} - \bm{X}_0\|_2$ follows the Hoyt distribution (also known as Nakagami-$q$ distribution) \cite{paris2009nakagami} with the pdf:
\begin{align}
    f(x; q, \Omega) &= \frac{1+q^2}{q \Omega} \, x\exp\left( -\frac{(1+q^2)^2 }{4q^2\Omega}x^2\right) I_0\left( \frac{1-q^4}{4q^2\Omega}x^2\right),  
\end{align}
where $\Omega$ is the expectation value of $ \|\bm{X} - \bm{X}_0\|_2$,
and $q$ is the shape parameter $(0<q<1)$ that depends on the variance of $(X-X_0)$ and $(Y-Y_0)$:
\begin{align}
    q &= \text{min} \left\{ \frac{\sigma_{s1}^2 + \sigma_{\mE_{\phi}}^2D^2}{\sigma_{s2}^2 + \sigma_{\mE_{\varphi}}^2D^2}, \frac{\sigma_{s2}^2 + \sigma_{\mE_{\varphi}}^2D^2}{\sigma_{s1}^2 + \sigma_{\mE_{\phi}}^2D^2}   \right\} \nonumber \\
    &= \text{min} \left\{\left(\frac{\cos(\phi)}{\cos(\varphi)}\right)^2,\left( \frac{\cos(\varphi)}{\cos(\phi)}\right)^2  \right\}. 
\end{align}
Therefore, in this case, the coverage probability $\p(\mathcal{C})$ is given by
\begin{align}
    \mathbbm{P}(\mathcal{C}) &\coloneqq \p(\{\| \bm{X} - \bm{X}_0 \|_2 < \rho_f -\rho_{\textsf{uav}} \}) = \int_{0}^{\rho_f - \rho_{\textsf{uav}}} f(x; q, \Omega) \, dx  \nonumber \\
    &= \int_{0}^{\rho_f - \rho_{\textsf{uav}}} \frac{1+q^2}{q \Omega} \, x\exp\left( -\frac{(1+q^2)^2 }{4q^2\Omega}x^2\right) I_0\left( \frac{1-q^4} {4q^2\Omega}x^2\right)\,dx.
    \label{hard}
\end{align}
Since the integral in (\ref{hard}) cannot be derived in closed-form, we will have to evaluate it numerically.



\subsection{Derivation of $\mathbbm{P}(\mathcal{D}|\mathcal{C})$}
As discussed before, $\mathbbm{P}(\mathcal{D} |\mathcal{C})$ is the average detection probability given that the beam footprint completely covers the UAV aperture. Here, let us assume that the beam footprint is centered at origin, the UAV receiver's lens center lies at the point $\bm{x}_0$, and that the beam footprint $\rho_f$ is greater than the UAV receiver lens radius $\rho_{\textsf{uav}}$.   Then,  the Euclidean distance between the beam center and the lens center is $\| \bm{x}_0 \|_2$. In this case, the received or captured signal energy by the UAV from the FSO Tx---given that the Tx beam footprint completely covers the receiver lens---is: 
\begin{align}
    E_{\textsf{uav}} \coloneqq (1-\alpha)E_t\iint_{A\left(\bm{x_0}\right)} g(x, y)\, dy \, dx \cdot \mathbbm{1}_{[0, \, \rho_f - \rho_{\textsf{uav}})}\left(\| \bm{x}_0 \|_2 \right),\label{power1}
\end{align}
where $(1-\alpha)E_t$ is the total transmitted power by the FSO Tx and $g(x,y) \coloneqq  \frac{1}{2 \pi \rho_f^2}\exp(- \frac{(x^2+y^2)}{2 \rho_f^2})$. The function $\mathbbm{1}_A(x)$ is an \emph{indicator} function that is one when $x \in A$ for a measurable set $A$, and is zero otherwise. Here, we denote $A(\bm{x}_0)$ as the region of the lens centered at point $\bm{x}_0$. The integral in \eqref{power1} cannot be computed in closed-form. Thus, to compute a closed-form expression, two approximations have to be used.

\begin{figure}
    \centering
    \subfigure[$\rho_{\textsf{uav}}<\rho_f$]{
    \label{1 Center}
    \includegraphics[width=0.3\textwidth]{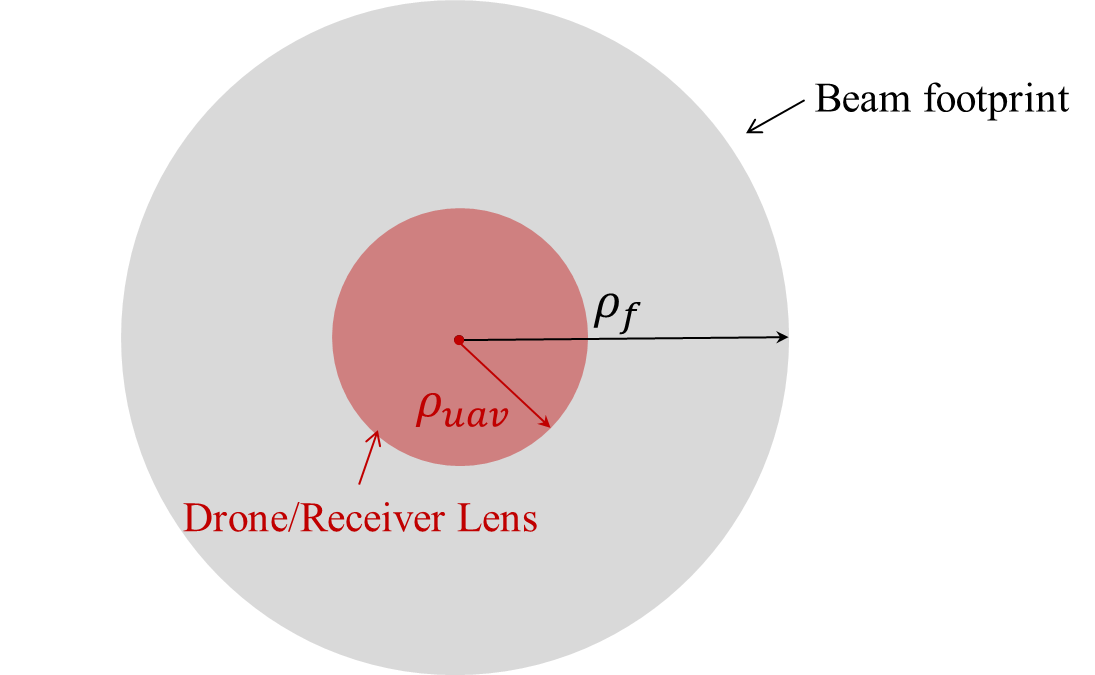}}
    \hspace{8mm}
    \subfigure[$\rho_{\textsf{uav}}\ll\rho_f$]{
    \label{2 Center}
    \includegraphics[width=0.3\textwidth]{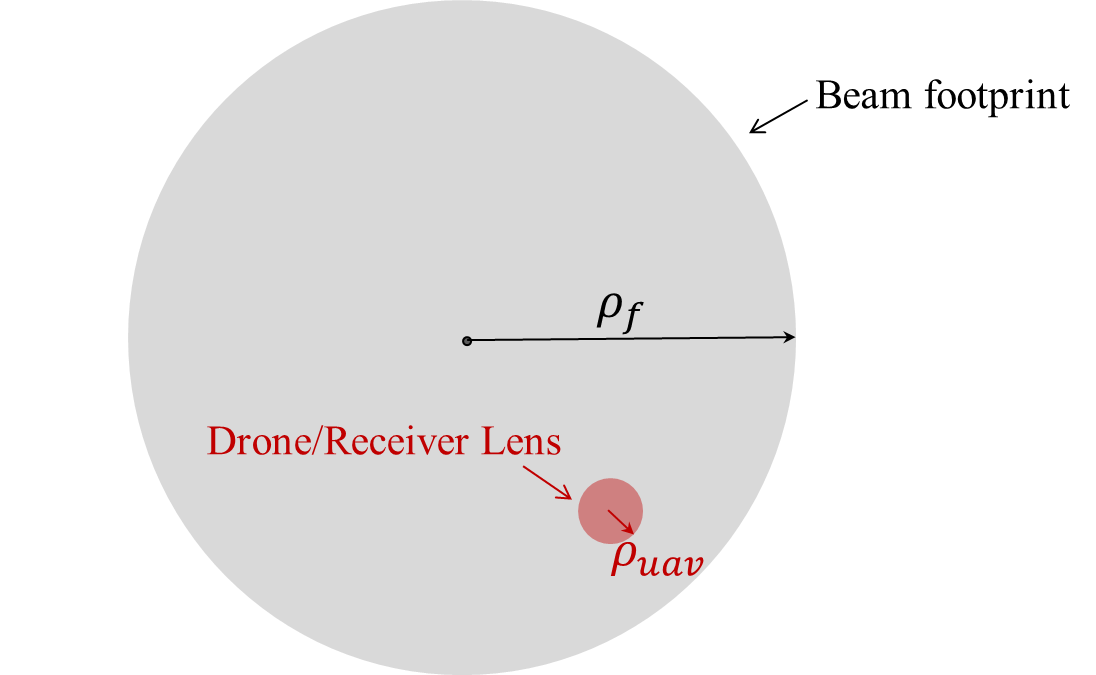}}
    \caption{The left figure represents the case when the center of the UAV and beam footprint coincide. The figure on the right  represents the scenario when the UAV is regarded as a point detector.}
    \label{Center}
\end{figure}

\subsubsection{ \text{Beam Center and UAV Lens Center Coincide}}

In this case, the centers of the UAV lens and the beam footprint coincide, as shown in Fig. \ref{1 Center}. Here,  we can transform the integral into the polar coordinates. Then the (average) received energy of the receiver is given by 
\begin{align}
   E_{\textsf{uav}} &= (1-\alpha) \frac{E_t}{N_0}  \iint_{\mathcal{B}(\rho_{\textsf{uav}}) } \frac{1}{2\pi \rho_f^2} \exp\left(-\frac{x^2+y^2}{2\rho_f^2}\right)dx\,dy = \frac{(1-\alpha) \frac{E_t}{N_0} }{2\pi \rho_f^2} \int_0^{\rho_{\textsf{uav}}} \int_0^{2\pi}  \rho \exp\left(-\frac{\rho^2}{2\rho_f^2}\right)d\rho \,d\theta \nonumber \\
   &= (1-\alpha) \frac{E_t}{N_0} \left( 1 - \exp \left( - \frac{\rho_{\textsf{uav}}^2}{2 \rho_f^2} \right) \right), \label{averaged power1}
\end{align}
where $\mathcal{B}(\rho_0)$ is an open ball of radius $\rho_0$ located at the origin.

\subsubsection{Point Detector Approximation}
When the distance between ground station and the UAV is large, we can assume $\rho_{\textsf{uav}} \ll \rho_f$. In this case, we treat the UAV receiver as a point detector, as shown in Fig. \ref{2 Center}. Then, the received energy at the location $(x,y)$ is given by $E_r(x,y) =(1-\alpha) E_t  \frac{1}{2 \pi \rho_f^2} \exp\left(- \frac{(x^2+y^2)}{2 \rho_f^2} \right) \pi \rho_{\textsf{uav}}^2$. Here, $(x, y)$ follows a truncated Gaussian distribution since we only considering the detection event after the coverage occurs. In this case, 
\begin{align}
    f(x, y) = \frac{1}{a_0} \left( \frac{1}{2 \pi (\sigma_s^2+\sigma_\mE^2 D^2)} \exp \left(- \frac{(x^2 + y^2)}{2 (\sigma_s^2+\sigma_\mE^2 D^2)} \right) \right) \cdot \mathbbm{1}_{\mathcal{B}(\rho_f)} (x, y),
\end{align}
where the constant $a_0$ is defined as the probability that the center of the beam and the UAV are within a distance of $\rho_f$ of each other. Let us define $\mathcal{B}(\rho_f)$ as an open ball of radius $\rho_f$ centered at the origin. We then have that the constant $a_0$ is defined as
\begin{small}
\begin{align}
    a_0 &\coloneqq \iint_{\mathcal{B}(\rho_f)} \frac{1}{2 \pi (\sigma_s^2+\sigma_\mE^2 D^2)} \exp \left(- \frac{(x^2 + y^2)}{2 (\sigma_s^2+\sigma_\mE^2 D^2)} \right) \, dy \, dx = \! \int_0^{2\pi} \!\! \! \int_0^{\rho_f} \!\! \frac{\rho}{2 \pi (\sigma_s^2+\sigma_\mE^2 D^2)} \exp \left(\!- \frac{\rho^2}{2 (\sigma_s^2+\sigma_\mE^2 D^2)} \!\right) \, d\rho \, d\theta \nonumber \\
    &= 1 - \exp \left( -\frac{\rho_f^2}{2(\sigma_s^2+\sigma_\mE^2 D^2)} \right).
\end{align}
\end{small}
In this case, the average received energy---given that the event $\mathcal{C}$ is true---is
\begin{small}
\begin{align}
\Euav  &=  \iint_{-\infty}^\infty E_r(x, y) f(x, y)\, dy \, dx = \iint_{\mathcal{B}(\rho_f)} E_r(x, y) f(x, y)\, dy \, dx \nonumber \\
&= (1-\alpha) \frac{E_t}{N_0} \iint_{\mathcal{B}(\rho_f)} \frac{1}{2 \pi \rho_f^2} \exp\left(- \frac{(x^2+y^2)}{2 \rho_f^2} \right) \pi \rho_{\textsf{uav}}^2 \left(\frac{1}{a_0} \frac{1}{2 \pi (\sigma_s^2+\sigma_\mE^2 D^2)} \exp \left(- \frac{(x^2 + y^2)}{2 (\sigma_s^2+\sigma_\mE^2 D^2)} \right) \right) \, dy \, dx \nonumber \\
&= \frac{(1-\alpha) \frac{E_t}{N_0} \rho_{\textsf{uav}}^2}{4 a_0 \pi\rho_f^2 \left( \sigma_s^2+\sigma_\mE^2D^2 \right)} \int_0^{2\pi}\int_0^{\rho_f} \rho \exp\left(- \frac{\rho^2}{2 \rho_f^2} \right) \exp \left(- \frac{\rho^2}{2 (\sigma_s^2+\sigma_\mE^2 D^2)} \right) \, dy \, dx  \nonumber \\
&= \frac{(1-\alpha) \frac{E_t}{N_0} \rho_{\textsf{uav}}^2}{4 a_0 \pi\rho_f^2 \left( \sigma_s^2+\sigma_\mE^2D^2 \right)} \cdot \frac{\pi \left(\sigma_s^2+\sigma_\mE^2D^2 \right)}{\sigma_s^2+\sigma_\mE^2D^2+\rho_f^2} \cdot 2\rho_f^2 \left(1- \exp\left( - \frac{( \sigma_s^2+\sigma_\mE^2D^2 + \rho_f^2)}{2  (\sigma_s^2 + \sigma_\mE^2 D^2)}  \right) \right)     \nonumber \\
&= \frac{(1-\alpha) \frac{E_t}{N_0} \rho_{\textsf{uav}}^2}{2 a_0 \left( \rho_f^2+\sigma_s^2+\sigma_\mE^2D^2 \right)} \left(1- \exp\left( - \frac{\sigma_s^2+\sigma_\mE^2D^2 + \rho_f^2}{2  (\sigma_s^2 + \sigma_\mE^2 D^2)}  \right) \right).
\end{align}
\end{small}


The energy $E_{\textsf{uav}}$ is converted into electric charge of magnitude $S_{\textsf{uav}} = \eta E_{\textsf{uav}}$ where $\eta$ is the photoconversion efficiency of the detector. The received signal $Z$ in this case is 
\begin{align}
    Z \coloneqq S_{\textsf{uav}} + W ,
\end{align}
where $W \sim \mathcal{N}(0, \sigma_W^2)$. The quantity $\sigma_W$ is noise standard deviation at the UAV receiver. Here, we compare $Z$ with a threshold $\Upsilon_0$ in order to decide between the two hypotheses: $H_0$ is that a signal pulse was not detected and $H_1$ is that a signal pulse was detected. Specifically,
\begin{align}
    Z \overset{H_0} {\underset{H_1}{\lessgtr}} \Upsilon_0.
\end{align}
Then the probability that $H_1$ is true (signal pulse is detected) is given by
\begin{align}
    \mathbbm{P}(\mathcal{D}|\mathcal{C}) &= \p\left(N > \Upsilon_0 - S_{\textsf{uav}}\right) = 1 -\Phi \left(\frac{\Upsilon_0 - S_{\textsf{uav}}}{\sigma_W}\right)
    \nonumber \\
    &= Q \left( \frac{\Upsilon_0 - S_{\textsf{uav}}}{\sigma_W} \right). \label{p_D }
\end{align}

Therefore, by substituting expression \eqref{p_C}, \eqref{hard} and \eqref{p_D } into the expression \eqref{P_N}, we have that the final expression of $p_N$.


\section{ Optimization Problem}\label{V}

\subsection{Why Do We Need to Optimize $\alpha$? } \label{Theoretical analysis}
When the energy split factor $\alpha$---where $0 < \alpha < 1$---is high or approaches 1, most of the energy will be allocated to the lidar. In this case, the lidar will provide a better estimate of the initial location of the UAV, which will cause the uncertainty sphere to diminish. However, the remaining energy dedicated for the FSO acquisition system will shrink, thereby leading to smaller energy per firing pulse. This results in a low probability of UAV detection. Thus, the number of times we update $(u_0, v_0)$ will increase, i.e., the number of acquisition attempts $X$---on average---will become large, thereby increasing the average total acquisition time $\E[T]$. However, when $\alpha$ is low or approaches 0, the energy used for the lidar system is quite small which causes the uncertainty sphere to become significantly large. In this scenario, the larger fraction of energy reserved for FSO acquisition system will result a higher amount of energy for each firing pulse. This will in turn increase the probability of detection $\mathbbm{P}(\mathcal{D}|\mathcal{C})$. However, the probability of coverage $\mathbbm{P}(\mathcal{C})$ will shrink with a large uncertainty sphere (please see \eqref{p_C}) which will lead to a smaller value of $p_N$ through the relationship \eqref{P_N}. In this scenario, we will exhaust all $N_0$ pulses to find the UAV, and with large probability, we will still not be able to locate the UAV. Thus, for a small $\alpha$, we will have to update $(u_0, v_0)$ several times which will again lead to a large value of $X$. Therefore, an optimal value of $\alpha$---that will minimize the average acquisition time---exists since both a high and a low $\alpha$ will considerably increase the expectation of the total acquisition time.

In order to quantify the acquisition time, we derive the average value and the CDF of acquisition time.







\subsection{Expectation of $T$}
From $\eqref{total acquisition time}$, the expectation of $T$ is given by
\begin{align}
    \E[T] & = (T_1+N_0 T_2) \,\E[X] + T_2 \,\E[N] - N_0 T_2 \nonumber \\
    & =  (T_1+N_0 T_2)\cdot \frac{1}{1 - (1-p_N)^{N_0}} +  \frac{T_2 \cdot [1-(1-p_N)^{N_0} \cdot (1+N_0 p_N)]}{[1-p_N-(1-p_N)^{N_0}] \cdot p_N} - N_0 T_2,
    \label{Expectation}
\end{align}
where $p_N$ is given by $\eqref{P_N}$ and $F_{N'}(x) = 1-(1-p_N)^{\lfloor x \rfloor}$ for $x \geq 0$.



\subsection{CDF of $T$}
From $\eqref{total acquisition time}$, the CDF (Cumulative distribution function) of $T$ is given by
\begin{align}
    \p(\{T\leq t\}) &= \p(\{ (T_1+N_0 T_2)X + T_2N - N_0 T_2 \leq t \}) \nonumber\\
&= \p\left(\left\{ \left(\frac{T_1}{T_2}+N_0 \right)X + N \leq \frac{t}{T_2}+N_0 \right\} \right).
\end{align}
Let $M = \frac{T_1}{T_2}+N_0 > N_0$, and $z = \frac{t}{T_2} + N_0$, then the CDF of T can be rewritten as:
\begin{align}
    \p(\{T\leq t\}) = \p(\{ M  X + N \leq z \}) = \p(\{ Y + N \leq z \}).
    \label{last1}
\end{align}
We can regard $MX$ as a function of a new random variable $Y = MX$, where the PMF of random variable $Y$ is given by:
\begin{align}
    \p(\{Y = y\}) = \p(\{X = \frac{y}{M} \}) \quad y = M, 2M, \dotsc
    \label{last2}
\end{align}

We now assume $Z$ as a new random variable: $Z \coloneqq  Y + N$. The CDF of the total acquisition time is given by:
\begin{align}
    \p(\{T\leq t\}) = \p(\{Z\leq z\}) 
    &=  \begin{cases}
    \frac{[p_X-1+(1-p_X)^{k}][(1-p_N)^{N_0}-1]}{(1-p_X)[1-p_N-(1-p_N)^{N_0}]} + \\
    \frac{p_X(1-p_X)^{k-1}[1-(1-p_N)^{z-kM}]}{1-p_N-(1-p_N)^{N_0}}, & z \in [kM + 1, kM + N_0] \cr
    \frac{[1-(1-p_X)^{k}][(1-p_N)^{N_0}-1]}{1-p_N-(1-p_N)^{N_0}}, &z \in [kM + N_0 + 1, (k+1)M] 
    \end{cases}
    \label{cao}
\end{align}
where $k = \lfloor \frac{z-1}{M} \rfloor$, $M = \frac{T_1}{T_2}+N_0$, $z = \frac{t}{T_2} + N_0$. The derivation of (\ref{cao}) is produced in the appendix.

\subsection{Optimization Problem}\label{Op problem}
Based on the discussion above, we propose three optimization problems. The objective functions we want to optimize are the expectation and CDF of the total acquisition time $T$ given by (\ref{Expectation}) and (\ref{cao}), respectively.

\subsubsection{Optimization of $\alpha$ for $\E[T]$}
\label{opt_prob_1}
\begin{equation}
\begin{aligned}
 \underset{\alpha }{\text{minimize }}
& \quad   \E[T]   \\
 \text{subject to } \quad  & i)\,  N_0 = n_0, \\ 
 & ii)\, 0<\alpha<1.
 \label{optimization 1}
\end{aligned} 
\end{equation}

\subsubsection{Optimization of $N_0$ for $\E[T]$}
\label{opt_prob_2}
\begin{equation}
\begin{aligned}
 \underset{N_0 }{\text{minimize }}
& \quad   \E[T]   \\
 \text{subject to } \quad  & i)\,  \alpha = \alpha_0, \\
 & ii)\, n_0<N_0<n_1
 \label{optimization 2}
\end{aligned} 
\end{equation}

\subsubsection{Optimization of $\alpha$ for CDF}
\label{opt_prob_3}
\begin{equation}
\begin{aligned}
 \underset{\alpha }{\text{maximize }}
& \quad   \p(\{T\leq t\})   \\
 \text{subject to } \quad  & i)\, N_0 = n_0, \\
 & ii)\, 0<\alpha<1.
 \label{optimization 3}
\end{aligned} 
\end{equation}

\section{Simulation Results and Commentary}\label{VI}
In this section, we summarise the simulation results of the optimization problems
mentioned in Section \ref{Op problem}. The set of the default parameter values of the experimental setup is shown in TABLE I. 

{ For our simulations, we have chosen (approximately) 50 mrads as the half angle beamwidth of lidar. This beamwidth is chosen by taking into account the height and speed of the UAV. The minimum distance between the lidar and the UAV is the height of the UAV from ground. In our simulations, the height of the UAV from the ground is chosen to be 100 m. Therefore, the (minimum) footprint/radius of the lidar beam at the UAV is $50 \times 10^{-3} \times 100 = 5$ m.  The minimum speed of most military fixed-wing UAVs is approximately 40 km/h (which translates to approximately 11 m/s) \cite{UAV_speed}. Assuming that the lidar fires a pulse once every 100 ms, the UAV only moves a distance of approximately 1 m between each pulse time. Furthermore, let us assume that the uncertainty sphere has a radius of 4 m. In this scenario, a lidar beam of 5 m is approximately sufficient to update the location $(u_0[n], v_0[n])\footnote{The index $n$ represents a discrete-time instant.}$ of the UAV based on the return signal even if the lidar is pointing at the previous location estimate $(u_0[n-1], v_0[n-1])$. This is true because the lidar beam radius of 5 m is approximately sufficient to cover uncertainty error of 4 m and the UAV movement of 1 m between each lidar pulse. }

\begin{table}[ht]
    \centering
    \begin{center}
    \scalebox{1}{
    \begin{tabular}{|{c}|{c}|{c}|}
         \hline
         \hline
         \textbf{Symbol} & \textbf{Parameter} & \textbf{Simulation Value} \\ \hline
         $E_t$ & Total transmit energy (Joule) & 10 \\ \hline
         $D$ & Distance between lidar and ground station (m) & 100 \\ \hline
         $\theta_l$ & Lidar beam width (half angle) (radians) & { 0.05} \\ \hline
         $\rho_l$ & Lidar beam radius at $D$ (m) & 10 \\ \hline
         $\theta_f$ & Tx beam width (half angle) & $5\times10^{-4}$ \\ \hline
         $\rho_f$ & Tx beam radius at $D$ (m) & 0.5 \\ \hline
         $a_l$ & Radius of of the lidar telescope (m) & 0.5 \\ \hline
         $\sigma$ & Radar cross-section of the UAV & 0.2 \\ \hline
         $\rho_{\textsf{uav}}$ & UAV telescope radius (m) & 0.01 \\ \hline
         $\phi$ & Elevation angle & 0.1 \\ \hline
         $\varphi$ & Azimuth angle & 0.6 \\ \hline
         $F$ & Focal length of the lidar receiver telescope lens & $1\times10^{-3}$ \\ \hline
         $N_0$ & Maximum number of firing pulses & 10  \\ \hline
         $\eta$ & Photon conversion efficiency factor & 0.5  \\ \hline
         $\sigma_N$ & Noise standard deviation at the UAV receiver & $1\times10^{-5}$ \\ \hline
         $T_1$ & Round-trip time of the pulse and process ($s$) & $1\times10^{-3}$ \\ \hline
         $T_2$ & Time interval between each pulse of the shotgun technique ($s$) & $1\times10^{-3}$ \\ \hline
         $t$ & Threshold ($s$) & $12$ \\ \hline
         \hline
    \end{tabular}
    }
    \hspace{10mm}
    \caption{\textbf{Table Of simulation parameters}}
    \end{center}
    \label{table}
\end{table}
\subsection{Optimization for $\E[T]$}\label{match1}
In this section, we will use $\E[T]$ in (\ref{Expectation}) as the objective function to obtain the optimal value of $\alpha$ and $N_0$. We have considered two scenarios of the uncertainty sphere: the circular sphere and the elliptical sphere. Additionally,  we simulate the cases when the centers of beam footprint and UAV aperture coincide, and when $\rho_{\textsf{uav}}\ll \rho_f$ (point receiver approximation).
\subsubsection{Circular Sphere}\label{match2}

\begin{figure}[ht]
    \centering
    \subfigure[Centers of beam and UAV aperture coincide ]{
    \label{Fig_a_a}
    \includegraphics[width=0.4\textwidth]{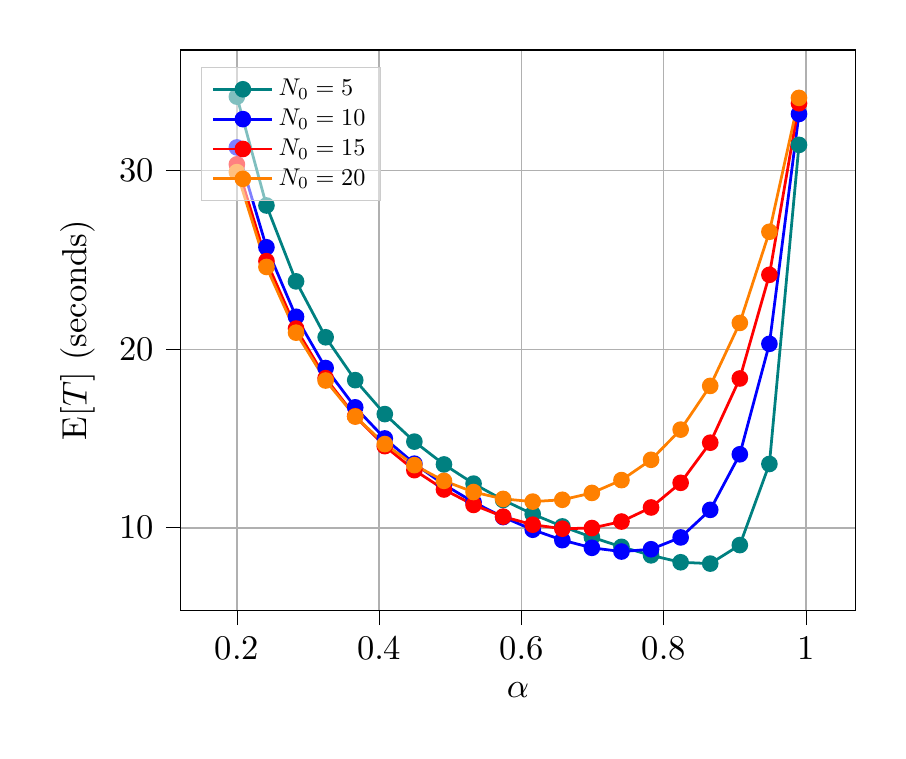}}
    \hspace{8mm}
    \subfigure[Point receiver approximation ]{
    \label{Fig_a_b}
    \includegraphics[width=0.4\textwidth]{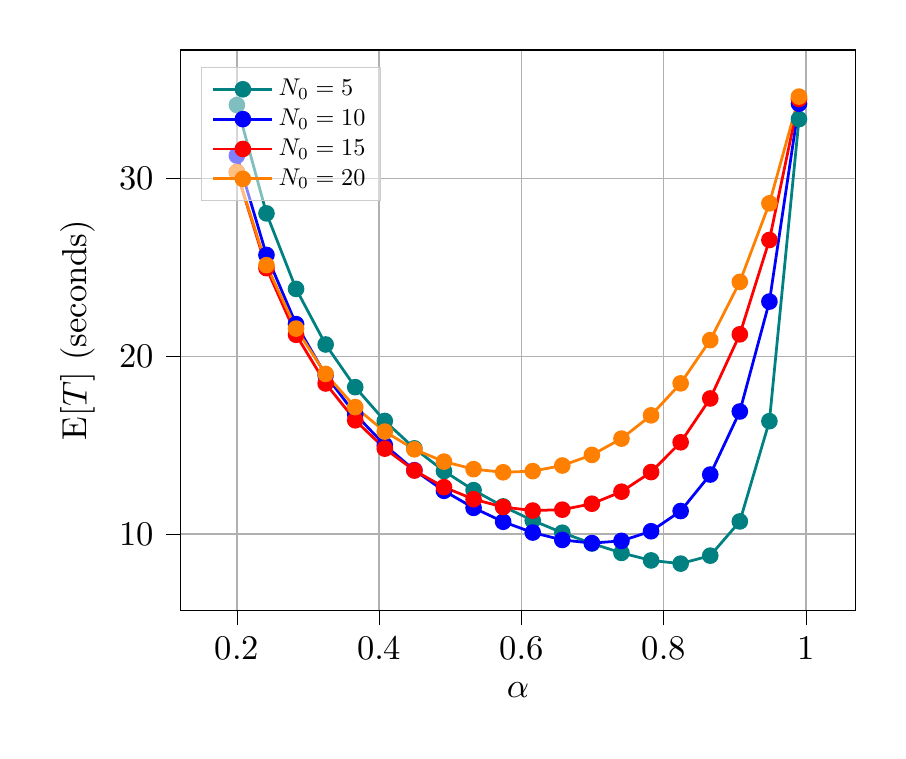}}
    \caption{The figure  shows the average total acquisition time $\E[T]$ as a function of energy split factor $\alpha$ for different $N_0$ (maximum number of firing pulses).  For both cases, the shape of the uncertainty sphere is circular symmetric.}
    \label{Fig1_alpha}
\end{figure}

Fig. \ref{Fig1_alpha} indicates the performance of our proposed system, that is, the average total acquisition time as a function of energy split factor $\alpha$ for four different values of $N_0$. As analyzed in (\ref{Theoretical analysis}), there exists an optimal value of $\alpha$ that provides the optimal performance of the proposed system. The left figure represents the case when the centers of beam and UAV aperture coincide, while the right figure represents the receiver is regarded as a point. We conclude from this figure that for both scenarios, the larger the number of firing pulses $N_0$, the smaller the optimal $\alpha$ that provides the best system performance (or the minimum of $\E[T]$). This is because as $N_0$ increases, the energy allocated to each pulse will decrease since the total energy for all pulses is fixed at $(1-\alpha)E_t$. Therefore, the probability that the UAV detects the pulse successfully $\p(\mathcal{D}|\mathcal{C})$ will diminish. Hence, in order to improve $\p(\mathcal{D}|\mathcal{C})$, we need to allocate more energy to the FSO acquisition system, i.e., decrease the energy split factor $\alpha$. Thus, the optimal value of $\alpha$ will decrease as $N_0$ increases.  Moreover, for the interval $0.4<\alpha<0.8$, there is an obvious intersection between the curves for $N_0=5,10$ and $N_0=15,20$, indicating that the performance of the acquisition system does not change monotonically with the decrease or increase of $N_0$. Therefore, it is only appropriate to optimize  $\E[T]$ with respect to $N_0$ which is the motivation for optimization Problem~2 in (\ref{optimization 2}).

\begin{figure}[ht]
    \centering
    \subfigure[Centers of beam and UAV aperture coincide ]{
    \label{Fig_N0_a}
    \includegraphics[width=0.4\textwidth]{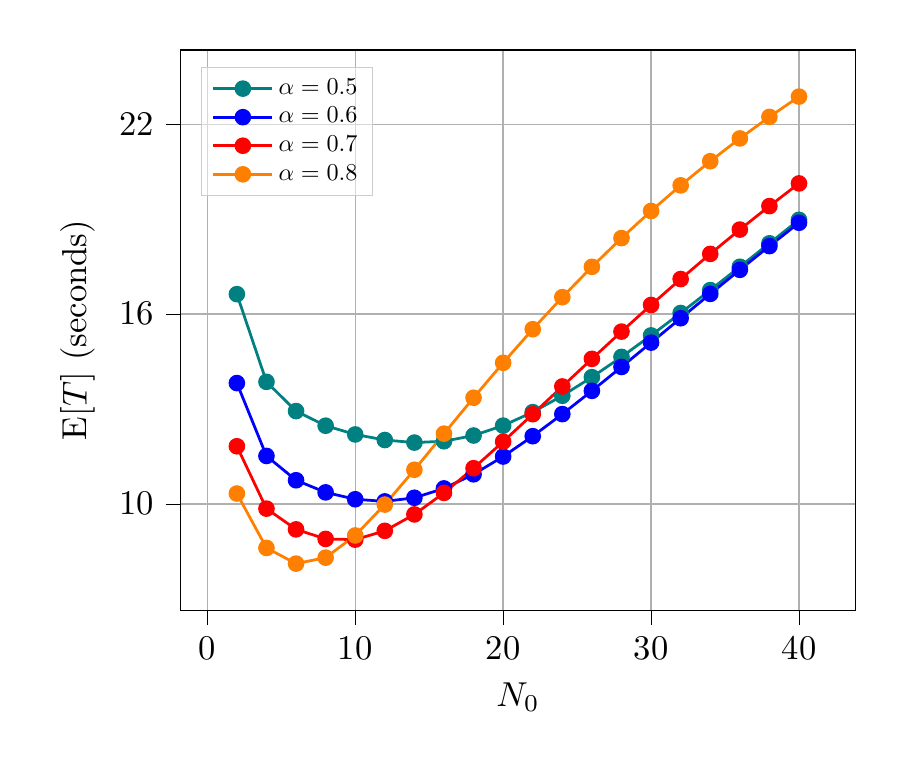}}
    \hspace{8mm}
    \subfigure[Point receiver approximation  ]{
    \label{Fig_N0_b}
    \includegraphics[width=0.4\textwidth]{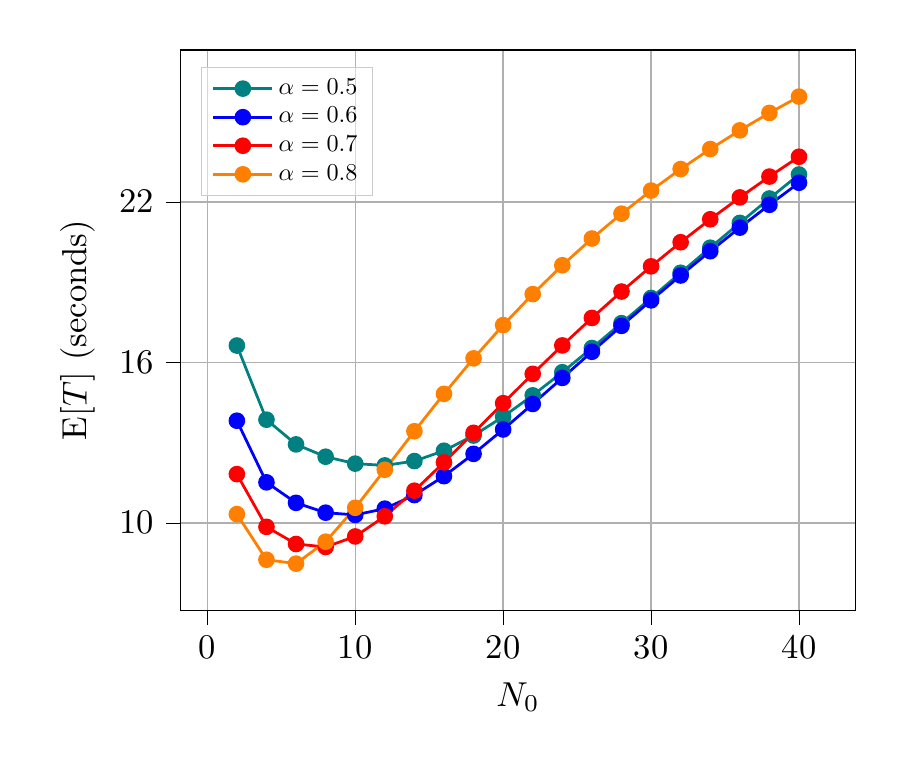}}
    \caption{The figure shows the average total acquisition time $\E[T]$ as a function of maximum number of firing pulse $N_0$ for different $\alpha$ (energy split factor). For both cases, the shape of the uncertainty sphere is circular symmetric.}
    \label{Fig2_N0}
\end{figure}

Fig. \ref{Fig2_N0} illustrates the average total acquisition time as a function of $N_0$ for four different values of $\alpha$. As expected, there is a particular value of $N_0$ that will minimize $\E[T]$. We can see from this figure that there is also an intersection between $\alpha = 0.5,0.6$ and $\alpha=0.7,0.8$, which corresponds to the convex nature of $\E[T]$ as a function of  $\alpha$. 

\subsubsection{Elliptical Sphere}
We now consider the more general scenario $\phi \neq \varphi$, i.e., when the variance of the elevation and azimuth angle is not the same. For simplicity, we will optimize $\alpha$ and $N_0$ with respect to $\E[T]$ only for the case of $\rho_{\textsf{uav}}\approx\rho_f$ (point receiver approximation) in this subsection.

\begin{figure}[ht]
    \centering
    \subfigure[Elliptical Sphere case with respect to $\alpha$ ]{
    \label{Ellptical_sub_a}
    \includegraphics[width=0.4\textwidth]{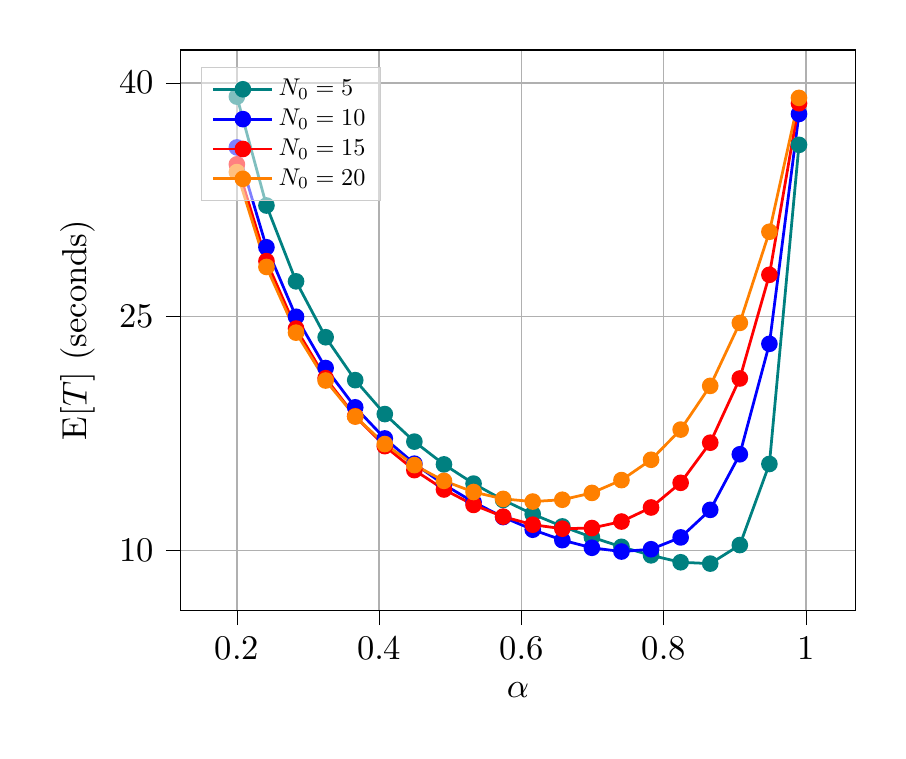}}
    \hspace{8mm}
    \subfigure[Elliptical Sphere case with respect to $N_0$ ]{
    \label{Ellptical_sub_b}
    \includegraphics[width=0.4\textwidth]{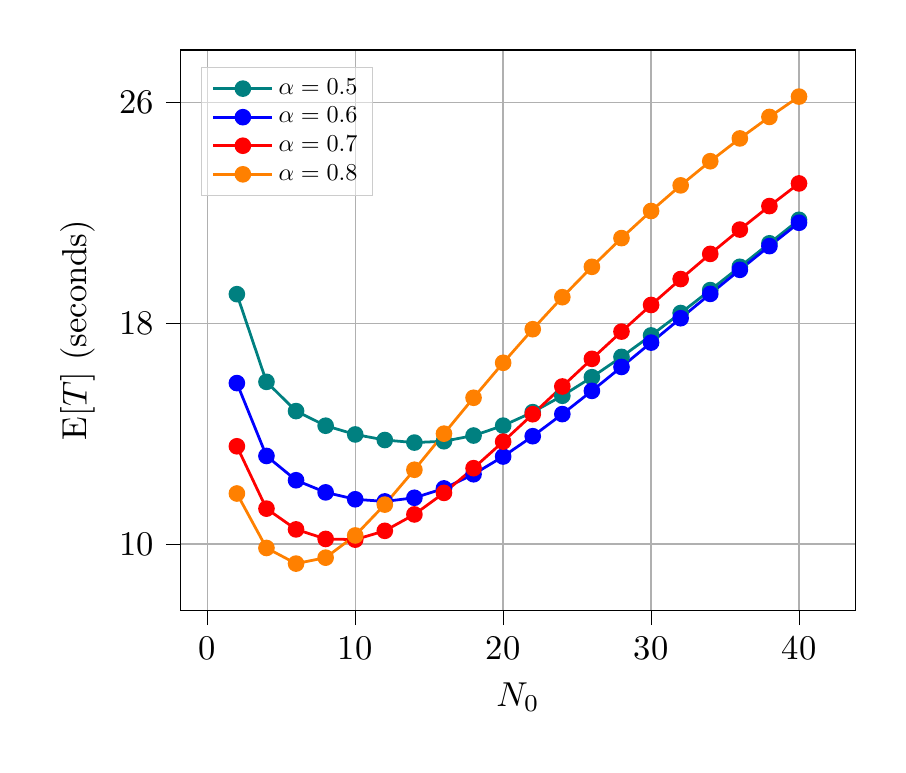}}
    \caption{The figure shows the average total acquisition time $\E[T]$ as a function of energy split factor $\alpha$ (left), and maximum number of firing pulse $N_0$ (right).}
    \label{Fig_ellptical}
\end{figure}

Fig. \ref{Fig_ellptical} depicts the average total acquisition time $\E[T]$ as a function of $\alpha$ and $N_0$ under the circumstance that the uncertainty sphere is elliptical. Compared to Fig.~\ref{Fig1_alpha}, this figure shows us a similar trend for the variation of $\E[T]$ as a function of both $\alpha$ and $N_0$. 


\begin{figure}
    \centering
    \includegraphics[width=0.4\textwidth]{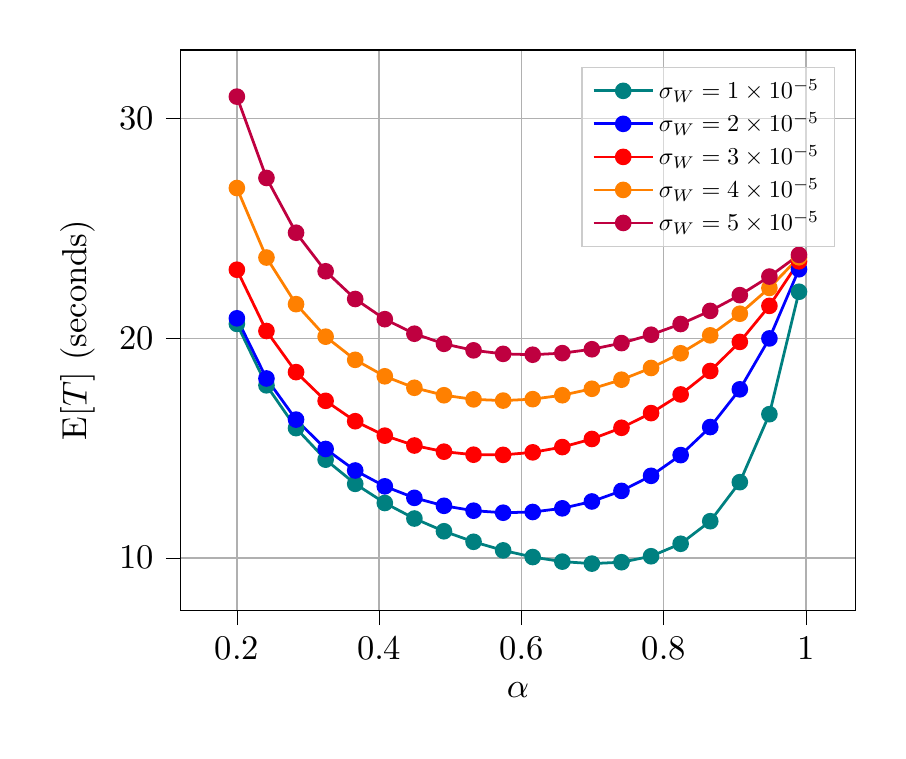}
    \caption{The average total acquisition time $\E[T]$ as a function of energy split factor $\alpha$ for different thermal noise power $\sigma_W^2$.}
    \label{ET_thermal}
\end{figure}

Fig. \ref{ET_thermal} demonstrates the average acquisition time $\E[T]$ as a function of $\alpha$ when the system suffers from different thermal noise power (denoted by $\sigma_W^2$) values at the UAV receiver. From this curve, we find that our proposed system's performance will deteriorate (the average acquisition time will increase)  with high noise power. We also note from this figure that the larger the noise power, the smaller the value of  optimal $\alpha$ (the larger the energy of power we have to allocate to FSO acquisition system to maximize the probability of detection).

\subsection{Optimization for CDF of $T$}
In this section, we use the cumulative distribution function (CDF) of the total acquisition time in (\ref{cao}) as the objective function to optimize energy split factor $\alpha$. Here, the system performance is measured by the probability that the total acquisition time is less than a given threshold $t$. In this case, the higher the CDF at some number $t$, the more improbable it is  that the acquisition time $T$ will exceed $t$, which indicates a superior system performance.

\begin{figure}
    \centering
    \subfigure[Circular sphere with respect to $\alpha$ ]{
    \label{CDF_case}
    \includegraphics[width=0.4\textwidth]{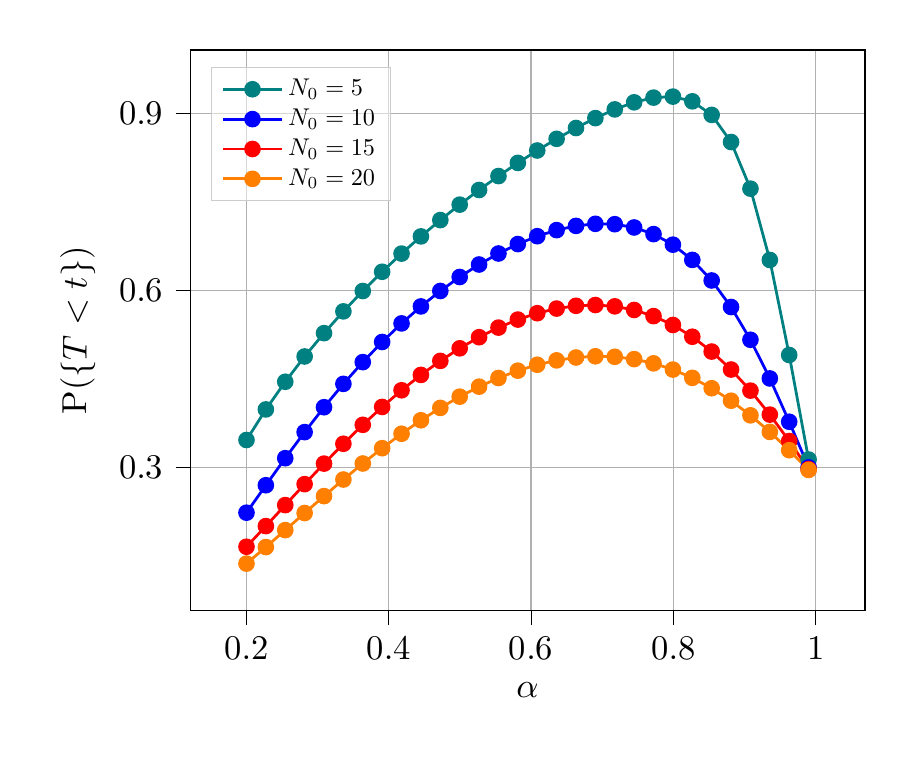}}
    \hspace{8mm}
    \subfigure[Circular sphere for different noise power $\sigma_W^2$ ]{
    \label{CDF_thermal}
    \includegraphics[width=0.4\textwidth]{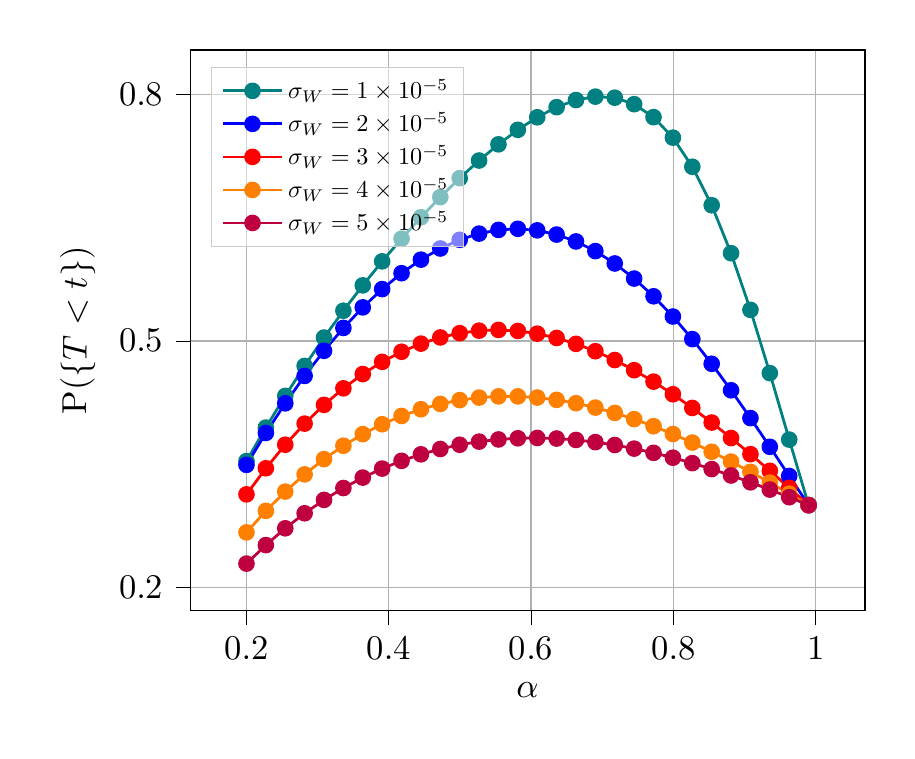}}
    \caption{The figure shows the probability of the total acqusition time below a threshold or the cumulative distribution function of the total acquisition time (point receiver approximation) for different $N_0$ (left) and thermal noise power $\sigma^2_W$ (right). The value of threshold $t=12$ seconds.}
    \label{Fig_CDF}
\end{figure}

Fig. \ref{Fig_CDF} shows the CDF $\p(\{ T \leq t\})$ as a function of $\alpha$ for  different values of firing pulses $N_0$ and thermal noise power $\sigma_W^2$.  We observe that there also exists an optimal energy split factor $\alpha$ that maximizes the CDF for different values of $N_0$.  As $N_0$ increases, the energy per pulse will shrink, and the probability of detection, $\p(\{ \mathcal{D} | \mathcal{C} \})$, will suffer. Therefore, as $N_0$ grows, the optimal $\alpha$ has to decrease in order to allocate more energy to the FSO acquisition system. In this figure, no intersection exists between the four curves, which demonstrates  the system performance will begin to degrade as $N_0$ increases beyond $5$. 

\begin{figure}
    \centering
    \includegraphics{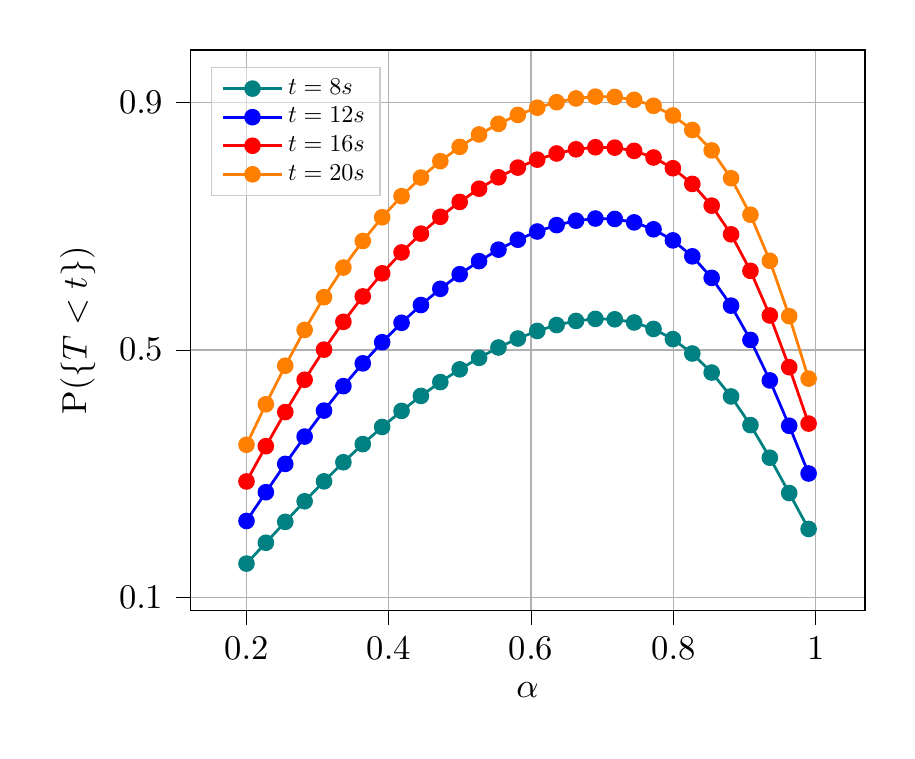}
    \caption{The cumulative distribution function of the total acquisition time (point approximation receiver) as a function of $\alpha$ for different thresholds $t$.}
    \label{threshold}
\end{figure}

Fig. \ref{threshold} presents the CDF $\p(\{ T \leq t\})$ as a function of $\alpha$ for different values of threshold $t$. From this figure, we find that the variation of threshold $t$ has no effect on the optimal value of $\alpha$.


\section{Conclusion}\label{VII}
In this paper, we have analyzed free-space optical communications that is assisted by a lidar in order to acquire a mobile airborne terminal with narrow laser beams. In this regard, we optimized the energy allocation between the lidar and narrow-beam FSO acquisition subsystems in order to minimize the average acquisition time and maximize the CDF of acquisition time. We considered the optimization problem for different scenarios of thermal noise power, shape of uncertainty sphere, and beam footprint.  Based on our results, we showed that there exists an optimal energy split factor $\alpha$ that can minimize the  acquisition time for the lidar-assisted acquisition system.  The analysis in this study can be applied directly to design efficient acquisition algorithms for lidar-assisted acquisition systems that are particularly suitable in GPS restricted environments. 

For future work, we will improve the lidar-assisted acquisition algorithm  presented in this study by the application of Bayesian filtering algorithms---such as Kalman and particle filtering---for efficient tracking of the UAV terminal with a lidar. This will lead to a smaller uncertainty sphere for the purpose of acquisition which will help us maximize the acquisition performance of our proposed system. 

\section{Appendix}
\subsection{Appendix~A}
Let $f(x) = \textsf{Ei}(x) - \ln(x) - \gamma$, where $x>0$, and then take the first derivative
\begin{align}
    f'(x) = \frac{e^x}{x} - \frac{1}{x} = \frac{e^x - 1}{x} \geq 1 \text{  for  } x>0,
\end{align}
this means that $f(x)$ is monotonically increasing in the domain, and then we take the limit when $x \rightarrow 0$
\begin{align}
    \underset{x\rightarrow 0}{\text{lim}}\, \frac{\textsf{Ei}(x)}{\text{ln}(x) - \gamma} = \underset{x\rightarrow 0}{\text{lim}}\, \frac{\frac{e^x}{x}}{\frac{1}{x}} = 1,
\end{align}
thus $f(x) > f(0) > 0$ for every $x>0$, and therefore $\left(-\gamma + \textsf{Ei}(\lambda_U) - \ln (\lambda_U)\right)$ is always positive.

\subsection{Appendix~B}
As for the derivation for the CDF of the total acquisition time $T$:

Based on \eqref{last1}, \eqref{last1},
let $Z = Y + N$, where the value range of $Z$ is $[M + 1, M + N_0], [2M + 1, 2M + N_0], \dotsc$ (since $N<N_0$), which can be written as $[kM + 1, kM + N_0], \text{where } k = 1, 2, \dotsc \in N^*$.

Then the pmf of $Z$ is given by: 
\begin{align}
    \p(\{Z = n\}) &= \p(\{Y = kM\}) \ \p(\{N = n-kM\}), \quad n \in [kM + 1, kM + N_0]  \nonumber \\
    &=  \p(\{X = k\}) \ \p(\{N = n-kM\}), \quad n \in [kM + 1, kM + N_0]   \nonumber \\
    &=  (1-p_X)^{k-1}p_X \frac{(1-p_N)^{n-kM-1}p_N}{1-p_N-(1-p_N)^{N_0}}, \quad n \in [kM + 1, kM + N_0]   \label{pmf of Z}
\end{align}
since $\p(\{Z = n\}) = 0$ when $n \in [kM+N_0+1,(k+1)M]$.

Here we have to specify the relationship between $k$ and $n$, i.e, $kM + 1 \leq n \leq kM + N_0 \implies k \leq \frac{n-1}{M} \leq k + \frac{N_0 - 1}{M} < k + 1$ since $M > N_0$, thus we obtain $k = \lfloor \frac{n-1}{M} \rfloor$.

As for the CDF of $Z$, we derived it step by step:

For $z \in [M+1,M + N_0]$, 
\begin{align}
    P(\{Z\leq z\}) &= \sum_{n=M+1}^{z}P(\{Z=n\}) = \sum_{n=M+1}^{z}P(\{Y = M \}) P(\{ N = n-M \}) \nonumber\\
    &= \sum_{n=M+1}^{z}P(\{X = 1 \}) P(\{ N = n-M \}),
\end{align}
when $z \in [M+ N_0 + 1,2M]$, 
\begin{align}
    P(\{Z\leq z\}) = \sum_{n=M+1}^{M + N_0}P(\{Z=n\}) = \sum_{n=M+1}^{M + N_0}P(\{X = 1 \}) P( \{ N = n-M \}),
\end{align}
when $z \in [2M+1,2M+N_0]$, 
\begin{align}
    P(\{Z\leq z\}) &= \sum_{n=M+1}^{M+N_0}P(\{Z=n\}) + \sum_{n=2M+1}^{z}P(\{Z=n\}) \nonumber\\
    &= \sum_{n=M+1}^{M+N_0}P(\{Y = M \}) P(\{ N = n-M \}) + \sum_{n=2M+1}^{z}P(\{Y = 2M \}) P(\{ N = n-2M \}) \nonumber \\
    &= \sum_{n=M+1}^{M+N_0}P(\{X = 1 \}) P(\{ N = n-M \}) + \sum_{n=2M+1}^{z}P(\{X = 2 \}) P(\{ N = n-2M \}),
\end{align}
when $z \in [2M+N_0+1,3M]$,
\begin{align}
    P(\{Z\leq z\}) &= \sum_{n=M+1}^{M+N_0}P(\{Z=n\}) + \sum_{n=2M+1}^{2M+N_0}P(\{Z=n\}) \nonumber\\
    &= \sum_{n=M+1}^{M+N_0}P(\{X = 1 \}) P(\{ N = n-M \}) + \sum_{n=2M+1}^{2M+N_0}P(\{X = 2 \}) P(\{ N = n-2M \}).
\end{align}
Based on mathematical induction, we have that for $z \in [kM + 1, kM + N_0]$,
\begin{align}
    \p(\{Z\leq z\}) &= \sum_{i=1}^{k - 1} \sum_{n=iM+1}^{iM+N_0}\p(\{Z=n\})  + \sum_{n =k M  + 1}^{z}\p(\{Z=n\}),
    \label{hohoh}
\end{align}
and for $z \in [kM + N_0 + 1, (k+1)M]$,
\begin{align}
    \p(\{Z\leq z\}) = \sum_{i=1}^{k} \sum_{n=iM+1}^{iM+N_0}\p(\{Z=n\}).
    \label{saddd}
\end{align}

By substituting \eqref{hohoh} and \eqref{saddd} into $\eqref{pmf of Z}$, we obtain the CDF of $Z$,
\begin{align}
    \p(\{Z\leq z\}) = \begin{cases} 
    \sum_{i=1}^{k - 1} \sum_{n=iM+1}^{iM+N_0}\frac{(1-p_X)^{i-1}p_X (1-p_N)^{n-iM-1}p_N}{1-p_N-(1-p_N)^{N_0}} + \\ \sum_{n=kM + 1}^{z}\frac{(1-p_X)^{k-1}p_X (1-p_N)^{n-kM-1}p_N}{1-p_N-(1-p_N)^{N_0}}, &   z \in [kM + 1, kM + N_0]     \cr \sum_{i=1}^{k} \sum_{n=iM+1}^{iM+N_0}\frac{(1-p_X)^{i-1}p_X (1-p_N)^{n-iM-1}p_N}{1-p_N-(1-p_N)^{N_0}} & z \in [kM + N_0 + 1, (k+1)M] \end{cases} 
\end{align}
after calculation, we rewrite the CDF of $Z$ in closed form,
\begin{align}
    \p(\{Z\leq z\}) &= 
    \begin{cases}
    \frac{[p_X-1+(1-p_X)^{k}][(1-p_N)^{N_0}-1]}{(1-p_X)[1-p_N-(1-p_N)^{N_0}]} + \\
    \frac{p_X(1-p_X)^{k-1}[1-(1-p_N)^{z-kM}]}{1-p_N-(1-p_N)^{N_0}}, & z \in [kM + 1, kM + N_0] \cr
    \frac{[1-(1-p_X)^{k}][(1-p_N)^{N_0}-1]}{1-p_N-(1-p_N)^{N_0}}, &z \in [kM + N_0 + 1, (k+1)M] 
    \end{cases}
\end{align}
where $k = \lfloor \frac{z-1}{M} \rfloor$, $M = \frac{T_1}{T_2}+N_0$, $z = \frac{t}{T_2} + N_0$, and therefore $\p(\{T\leq t\}) = \p(\{Z\leq z\})$.

\bibliographystyle{IEEEtran}
\bibliography{ref.bib, refs.bib, refs1.bib, ref3.bib}

\end{document}